\documentclass[journal]{IEEEtran}
\usepackage{amssymb}
\usepackage{amsmath}
\usepackage{algorithmic}
\usepackage{algorithm}
\usepackage{array}
\usepackage{cite}
\usepackage{amsfonts} 
\usepackage{balance}
\usepackage{graphicx} 
\usepackage{multicol}
\usepackage[caption=false,font=normalsize,labelfont=sf,textfont=sf]{subfig}
\usepackage[colorlinks,
linkcolor=blue,     
anchorcolor=blue, 
citecolor=blue,     
]{hyperref}
\newtheorem{theorem}{Theorem}
\newtheorem{lemma}{Lemma}

\begin{document}
	
	\title{NOMA-Based Hybrid Satellite-UAV-Terrestrial Networks for Beyond 5G Maritime \\Internet of Things}

	\author{Xinran~Fang, 
		Wei~Feng,~\IEEEmembership{Senior Member,~IEEE,} Yanmin~Wang,
		Yunfei~Chen,~\IEEEmembership{Senior Member,~IEEE,}\\ 
		Ning~Ge,~\IEEEmembership{Member,~IEEE}, and Zhiguo Ding,~\IEEEmembership{Fellow,~IEEE}
		\thanks{X.~Fang, W.~Feng, and N. Ge are with the Department of Electronic Engineering, Beijing National Research Center for Information Science and Technology, Tsinghua University, Beijing 100084, China (e-mail: {fxr20}@mails.tsinghua.edu.cn, {fengwei}@tsinghua.edu.cn, {gening}@tsinghua.edu.cn). 
			
			Y.~Wang is with the China Academy of Electronics and Information Technology, Beijing 100041, China (e-mail: yanmin-226@163.com). 
			
			Y.~Chen is with the School of Engineering, University of Warwick, Coventry CV4 7AL, U.K. (e-mail: {Yunfei.Chen}@warwick.ac.uk).
		
			Z.~Ding is with the School of Electrical and Electronic Engineering, The University of Manchester, Manchester M13 9PL, U.K. (e-mail: zhiguo.ding@manchester.ac.uk).}

	}

	\maketitle	
	\begin{abstract}
	Current fifth-generation (5G) networks do not cover maritime areas, causing difficulties in developing maritime Internet of Things (IoT). To tackle this problem, we establish a nearshore network by collaboratively using on-shore terrestrial base stations (TBSs) and tethered unmanned aerial vehicles (UAVs). These TBSs and UAVs form virtual clusters in a user-centric manner. Within each virtual cluster, non-orthogonal multiple access (NOMA) is adopted for agilely including various maritime IoT devices, which are usually sparsely distributed on the vast ocean. The nearshore network also shares spectrum with marine satellites. In such a NOMA-based hybrid satellite-UAV-terrestrial network, interference among different network segments, different clusters, as well as different users occurs. We thereby formulate a joint power allocation problem to maximize the sum rate of the network. Different from existing studies, we use large-scale channel state information (CSI) only for optimization to reduce system overhead. The large-scale CSI is obtained by using the position information of maritime IoT devices. The problem is non-convex with intractable non-linear constraints. We tackle these difficulties by adopting the max-min optimization, auxiliary function method, and successive convex approximation technique. An iterative power allocation algorithm is accordingly proposed, which is shown effective for coverage enhancement by simulations. This shows the potential of NOMA-based hybrid satellite-UAV-terrestrial networks for maritime on-demand coverage.
	\end{abstract}
	
	
	\section{Introduction}
	Nowadays, the fast development of maritime economies fosters maritime Internet of Things (IoT), bringing huge demand on maritime communications. Although advanced fifth-generation (5G) techniques bring much convenience to ground networks, they still cannot support efficient coverage on the sea. At present, the maritime communication system mainly consists of two parts, terrestrial base stations (TBSs) and marine satellites. TBSs have limited coverage while satellites usually have low data rates and large transmission delays, which make the communication on the sea far from satisfying. How to construct an effective maritime network remains open.

	Generally speaking, satellites are widely accepted to be an indispensable part of global coverage, but it is more efficient to rely on advanced terrestrial techniques for nearshore areas \cite{n8}. It is reported that the most maritime communication demands happen within the exclusive economic zone \cite{n2}. In this sense, a wide-band nearshore network is expected. However, TBSs are mostly deployed on high lands to overcome the influence of the earth curvature \cite{n3}, where basic infrastructures such as optical fibers are hard to deploy. This leads to site scarcity problems of TBSs and relaying them only becomes more and more insufficient. To address these issues, unmanned aerial vehicles (UAVs) could join in as a part of communication facilities \cite{n5}. Their easy-deployment merits could help alleviate the site problem of TBSs. Due to the generally higher altitude, UAVs can also provide wider coverage and better channels \cite{n4,nn4}. In a word, the coordination of TBSs and UAVs deserves to be well-designed, so as to extend the efficient terrestrial services farther from the shore.

	In addition, to accommodate the wide sparsity of maritime IoT \cite{n6} and agilely serve different users, the non-orthogonal multiple access (NOMA) transmission scheme can be considered \cite{wen1, wen2}. On the one hand, this scheme is designed to serve a group of users within one resource block, thus more connections and higher spectrum efficiency can be achieved compared with the orthogonal multiple access (OMA) method \cite{ref4,wen3}. On the other hand, NOMA could implement the successive interference cancellation (SIC) to extract far users' interference from near users' signals \cite{n7}. Far users can be better attended without much compromise of near users.

	Considering aggravated spectrum scarcity problems, nearshore networks may share spectrum with marine satellites. With the aid of NOMA, we believe such a NOMA-based hybrid satellite-UAV-terrestrial network is promising to bring new opportunities on the sea~\cite{n5}. Nevertheless, 
	it also faces unique challenges. Due to the geographical limitation of both deploying TBSs and dispatching tethered UAVs, the nearshore network is irregular, leading to complicated co-channel interference. In addition, the nearshore network and the satellite have mutual interference. In this work, we tackle these problems, so as to show the potential gain of NOMA-based hybrid satellite-UAV-terrestrial networks for maritime coverage enhancement. Particularly, we exploit the unique features of maritime communication channels and maritime users and accordingly design a joint resource allocation scheme to maximize the sum rate of the nearshore network with practical constraints including affordable system overhead and low computational complexity.

	\subsection{Related Works}
	To the best of our knowledge, few studies explored the potentials of UAVs on the sea. To improve the rate of a  maritime user, Li \emph{et al.}~\cite{n9} leveraged a fix-wing UAV and jointly optimized UAV’s trajectory and power. Wang \emph{et al.}~\cite{n10} extended the scenario to a multi-UAV case and proposed an on-demand coverage scheme. Zhang \emph{et al.}~\cite{n12} utilized the UAV as a relay and optimized its placement, so as to extend the nearshore coverage. In addition,  with only TBSs and satellites, Wei \emph{et al.}~\cite{n8} utilized maritime environment information to estimate channel states, which were further utilized to instruct the precoding design. Xiao \emph{et al.}~\cite{n11} devised a voyage-based cooperative resource allocation scheme, so as to reduce system power consumption. These studies provide valuable references for the improvement of hybrid maritime networks. But none of them takes the wide sparsity of maritime IoT into account and the corresponding coverage scheme is still missing.

	Beyond maritime settings, UAV-assisted hybrid networks have received intensive attention. As a representative example, Hua \emph{et al.}~\cite{n13}  considered a UAV-aided cognitive satellite-terrestrial system and jointly optimized the trajectory and transmit power of a UAV. Zhao \emph{et al.}~\cite{n15} applied the NOMA technique to support massive connections, and the bilateral interference between ground and UAV users is mitigated. Similarly, Peng \emph{et al.}~\cite{ad1} considered a UAV-assisted NOMA network and optimized the user scheduling, UAV's trajectory as well as NOMA precoding. Great potentials of UAVs in different communication scenarios have been well illustrated. However, in most practice, the UAV will not stand alone. The achievable performance of multi-UAV should be more emphasized. 

	When taking systematical designs into account, cooperative resource allocation is effective for interference mitigation. Liu \emph{et al.}~\cite{n16} proposed a process-oriented optimization framework and jointly allocated subchannels, transmit power, and hovering time of a UAV swarm. With the assistance of NOMA , Lin \emph{et al.}~\cite{n18} utilized the collaborative beamforming to tackle troubling interference and optimized power allocation for sum-rate maximization.
	Wang \emph{et al.}~\cite{n17} investigated a NOMA-based space-terrestrial network and maximized the system energy efficiency. These studies provide useful guidelines for interference mitigation, but most of them assume that full channel state information (CSI) can be acquired while ignoring the overhead problem. The corresponding cost remains to be reevaluated.

	To construct a wide-band nearshore network,  tethered UAVs can be applied for coverage enhancement. But few studies give detailed regimes on how to integrate UAVs into maritime networks. In addition, the wide sparsity of maritime IoT requires on-demand coverage. In this case, the performance of NOMA remains to be evaluated. Taking the harsh maritime environment into account, the acquisition of full CSI in the hybrid satellite-UAV-terrestrial network is challenging. It not only requires the interconnection of space, air, and ground components but also faces the problem of lagging feedback. In a nutshell, there still lack practical regimes for the on-demand coverage of maritime IoT in terms of a NOMA-based hybrid satellite-UAV-terrestrial network. 
	
	\subsection{Main Contributions}
	In this paper, we explore potential gains of the NOMA-based hybrid satellite-UAV-terrestrial network for maritime coverage enhancement, using unique features of maritime channels and maritime users. We summarize our main contributions as follows.

	\begin{enumerate}
	\item  We establish a NOMA-based hybrid satellite-UAV-terrestrial maritime network for on-demand coverage. Tethered UAVs are dispatched to form virtual clusters with TBSs, coordinately serving maritime IoT devices in a user-centric manner. The NOMA technique is adopted to agilely include various users, so as to accommodate the wide sparsity of maritime IoT. We also consider the spectrum sharing between nearshore clusters and maritime satellites, which mainly support far sea areas beyond the coverage of the nearshore network.
	
	\item
	We formulate an ergodic sum-rate maximization problem with the constraints of the inter-segment interference and the minimal quality-of-service (QoS) requirement.  To reduce system overhead, only large-scale CSI is used for optimization, which could be acquired based on position information. The problem is non-convex with intractable non-linear constraints. We first turn the maximization problem into a max-min problem and then construct a monotonous auxiliary function to help get rid of the intractable non-linear constraints. Based on the method of successive convex approximation, the problem is solved in an efficient iterative way.
	
	\item 
	We evaluate the proposed scheme by simulations. It shows that our power allocation scheme is effective to mitigate challenging inter-user, inter-cluster, and inter-segment interference and the proposed algorithm takes on low computational complexity. The potential gain of the NOMA-based hybrid satellite-UAV-terrestrial network is also shown. It is observed that nearshore networks can be enhanced by the designed virtual clusters, and the NOMA technique can accommodate the wide sparsity of maritime IoT.
	\end{enumerate}

	\subsection{Organization and Notation}
	The rest of this paper is organized as follows. Section \ref{section 2} introduces the system model of the maritime hybrid satellite-UAV-terrestrial network. The joint power allocation scheme is proposed in Section \ref{section 3} for maximizing the ergodic sum rate with practical constraints. Simulation results and conclusions are presented in Section \ref{section 4} and Section \ref{section 5}, respectively.
	
	Throughout this paper, vectors and matrices are represented by lower and upper boldface symbols. $\mathbb{C}^{M \times N}$ represents the set of $M\times N$ complex matrices and $\mathbf{I}_M$ is the unit matrix of $M\times M$. $||\cdot||$ denotes the Euclidean norm and $(\cdot)^H$ is the operation of transpose conjugate, $\mbox{det}(\cdot)$ represents the determinant operator. The complex Gaussian distribution with zero mean and $\sigma^2$ variance is denoted as $\mathcal{CN}(0,\sigma^2)$. $\mbox{E}_s (\cdot)$ is the expectation operation with respect to $s$.

	\section{System Model}
	\label{section 2}

	\begin{figure}[!t]
		\centering
		\includegraphics[width=3.5 in]{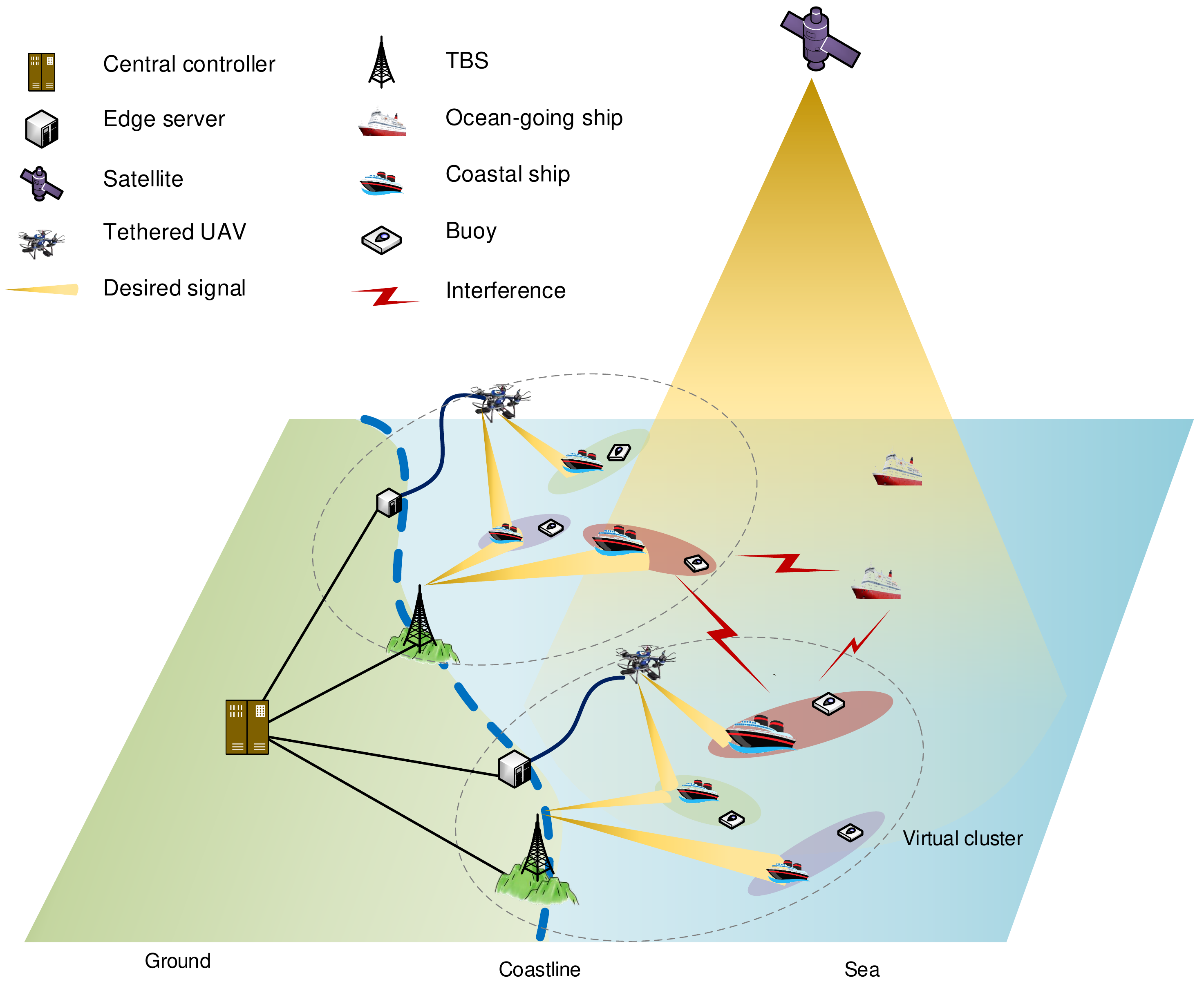}
		\caption{Illustration of a NOMA-based hybrid satellite-UAV-terrestrial network on the sea, where TBSs and UAVs form virtual clusters in a user-centric manner, and the coastal ship and the buoy are paired in each virtual cluster. The ocean-going ships are connected by the satellite. In this network, there exists challenging inter-user, inter-cluster, and inter-segment interference.}
		\label{Fig.1}
	\end{figure}

	As shown in Fig. \ref{Fig.1}, we consider a NOMA-based satellite-UAV-terrestrial network on the sea. In nearshore areas,  tethered UAVs are dynamically dispatched and form virtual clusters with TBSs for serving the users therein. 	
	When the virtual clusters are formed, we assume the positions of UAVs are fixed. These clusters can alleviate the site scarcity of TBSs and also extend their coverage scope. To accommodate the wide sparsity of maritime IoT, the NOMA scheme is applied, which provides great flexibility to pair different users. As an example shown in Fig. \ref{Fig.1}, NOMA is used to offer high-rate services to  coastal ships while carrying additional bits for buoys by serving them in pairs. Within a cluster, there are multiple users and different user pairs are allocated with orthogonal physical resource blocks (PRBs). Since the interference only occurs among the users using the same PRB, we thereby take one PRB and the corresponding users into account. As for the areas out of the nearshore network, the marine satellites provide complementary coverage. Inter-cluster connected edge servers are applied to collect position information and provide power supplies for UAVs. They are connected to the central controller, where the cooperative resource allocation is implemented.

	We assume there are $K$ virtual clusters and in cluster $k$ ($k=1,...,K$), two NOMA users equipped with $M_k$ antennas are served by $T_k$ TBSs and $U_k$ UAVs. All TBSs and UAVs are set to have one antenna. There are also $J$ single-antenna users served by the satellites. They are assumed to use the same frequency as considered NOMA users.

	We consider a composite channel model to depict both  small-scale and  large-scale fading. Denote $\mathbf{H}_m^{(k,i)}\in\mathbb{C}^{ M_k\times(T_i+U_i)}$ as the channel between user $m$ in cluster $k$ and transmitters in cluster $i$. It is given by
	\begin{equation}
	\mathbf{H}_m^{(k,i)}=\mathbf{S}_m^{(k,i)}(\mathbf{L}_m^{(k,i)})^\frac{1}{2}, \ m=1,2, \ k=1,...,K,
	\end{equation}
	where $\mathbf{S}_m^{(k,i)}\in\mathbb{C}^{M_k\times(T_i+U_i)}$ is the matrix of small-scale fading whose elements are independently and identically distributed (i.i.d)  Gaussian random variables of the $\mathcal{CN}(0,1)$ distribution. $\mathbf{L}_m^{(k,i)} \in \mathbb{C}^{(T_i+U_i)\times(T_i+U_i)}$ is the matrix of large-scale fading
	\begin{equation}
	\mathbf{L}_m^{(k,i)}=
	\left[
	\begin{array}{cccccc}
		l_{m,1}^{(k,i)} && {\huge\text{ o}}\\
		& \ddots & \\
		{\huge\text{ o}}&&l_{m,T_i+U_i}^{(k,i)}
	\end{array}
	\right],
	\end{equation}
	where $l_{m,n}^{(k,i)}, n=1,..,T_i, $ represents the large-scale fading of the TBS link and $l_{m,n}^{(k,i)}, n=T_i+1,..,T_i+U_i,$ represents the large-scale fading of the UAV link. For TBSs, we adopt the Hata propagation model~\cite{ref1}. Considering antenna gains, i.e, $g_t$, $l_{m,n}^{(k,i)}$ is calculated by
	\begin{equation}
	\label{equ1}
	\begin{split}
		l_{m,n}^{(k,i)}[\text{dB}]&=g_t[\text{dBi}]+(44.9-6.55\log_{10}h_t)\log_{10}\frac{d_{m,n}^{(k,i)}}{1000}\\
		&+(35.46-1.1h_r)\log_{10}f_c-13.82\log_{10}h_r \\
		&+0.7h_r+C+45.5,
	\end{split}
	\end{equation}
	where $h_r$(m) and $h_t$(m) represent the height of user terminals and TBSs, $d_{m,n}^{(k,i)}$(m) is the distance element, $f_c$(MHz) is the carrier frequency and $C$ is a constant related to the environment.  For UAVs, a model which considers both line-of-sight (LOS) and non-line-of-sight (NLOS) factors is adopted ~\cite{ref2}. With the antenna gains, i.e.,  $g_u$, $l_{m,n}^{(k,i)}$ is calculated by
	\begin{equation}
	\label{equ2}
	l_{m,n}^{(k,i)}[\text{dB}]=g_u[\text{dBi}]+\frac{A}{1+ae^{-b(\rho_{m,n}^{(k,i)}-a)}}+B_{m,n}^{(k,i)},
	\end{equation}
	where
	\begin{subequations}
	\begin{align}
		&A=\eta_{LOS}-\eta_{NLOS}, \\
		&B_{m,n}^{(k,i)}=20\log_{10}(d_{m,n}^{(k,i)})+20\log_{10}\big(\frac{4\pi f_c}{c}\big)+\eta_{NLOS}, \\
		&\rho_{m,n}^{(k,i)}=\frac{180}{\pi}\arcsin\big(\frac{h_u}{d_{m,n}^{(k,i)}}\big),
	\end{align}
	\end{subequations}
	and $h_u$(m) is the height of UAVs, $c$ is the speed of light, $\eta_{LOS}$, $\eta_{NLOS}$, $a$, $b$ are constants related to the environment. For further formulations, we transform $l_{m,n}^{(k,i)}[\text{dB}]$ into $l_{m,n}^{(k,i)}$, 
	\begin{equation}
	l_{m,n}^{(k,i)}=10^{-\frac{l_{m,n}^{(k,i)}[\text{dB}]}{10}},m=1,2,\ k,i=1,...,K.
	\end{equation}
	Similarly, the channel of satellite users is given by
	\begin{equation}
	\mathbf{h'}^{(j,k)}=\mathbf{s'}^{(j,k)}(\mathbf{L'}^{(j,k)})^\frac{1}{2}, j=1,...,J, \ k=1,...,K,
	\end{equation}
	where $\mathbf{s'}^{(j,k)} \in\mathbb{C}^{1\times (T_k+U_k)}$ denotes the small-scale fading and $\mathbf{L'}^{(j,k)} \in\mathbb{C}^{(T_k+U_k)\times (T_k+U_k)}$ is the large-scale fading, i.e,  $\mathbf{L'}^{(j,k)}=\text{diag}\big\{{l'}_1^{(j,k)},...,{l'}_{T_k+U_k}^{(j,k)}\big\}$.

	From the expression of \eqref{equ1} and \eqref{equ2}, it is easy to find that the large-scale fading is highly related to users' and transmitters' positions while the small-scale fading is unpredictable. To avoid the huge overhead, we assume only large-scale CSI is available. The corresponding position data could be obtained from  existing  monitoring systems such as the Automatic Identification System (AIS) \cite{ad4}.
	
	\section{Joint Power Allocation for Sum-Rate Maximization}
	\label{section 3}
	\subsection{Problem Formulation}
	Denote $\mathbf{x}^{(k)} \in\mathbb{C}^{(T_k+U_k)\times1}$ as the downlink signal in cluster $k $, and it is expressed as
	\begin{equation}
	\mathbf{x}^{(k)}= \mathbf{x}_1^{(k)}+ \mathbf{x}_2^{(k)}, \ k=1,...,	K,
	\end{equation}
	with $\mathbf{x}_1^{(k)}$, $\mathbf{x}_2^{(k)}$ representing the transmit signals of user 1 and user 2. The received signals can be given by
	\begin{subequations}
	\begin{equation}
		\mathbf{y}_1^{(k)}=\mathbf{H}_1^{(k,k)}\mathbf{x}_1^{(k)}+ \mathbf{H}_1^{(k,k)}\mathbf{x}_2^{(k)}+\sum_{i=1,i\not=k}^K\mathbf{H}_1^{(k,i)}\mathbf{x}^{(i)}+\mathbf{n}_1^{(k)},
	\end{equation}
	\begin{equation}
		\mathbf{y}_2^{(k)}=\mathbf{H}_2^{(k,k)}\mathbf{x}_2^{(k)}+\mathbf{H}_2^{(k,k)}\mathbf{x}_1^{(k)}+\sum_{i=1,i\not=k}^K\mathbf{H}_2^{(k,i)}\mathbf{x}^{(i)}+\mathbf{n}_2^{(k)},
	\end{equation}
	\end{subequations}
	where $\mathbf{n}_m^{(k)}\in\mathbb{C}^{M_k\times1} \ (m=1,2)$ represents the  additive white Gaussian noise of the $\mathcal{CN}(0,\sigma^2\mathbf{I}_{M_k})$ distribution. 

	Without loss of generality, we assume user 1 is the user who implements the SIC. When receiving the message, user 1 first decodes the content of user 2 and removes it from the received signal, subsequently decoding its own message. Considering the inter-cluster interference, the covariance matrix of interference plus noise of user 1 in cluster $k$ is given by 

	\begin{equation}
	\begin{aligned}
		\bar{\mathbf{Z}}_1^{(k)}&=
		\mathbf{E}_\mathcal{S}\big[\sum_{i=1,i\not=k}^K\mathbf{H}_1^{(k,i)}\mathbf{P}^{(i)}\mathbf{H}_1^{(k,i)^H}+\sigma^2\mathbf{I}_{M_k}\big]\\
		&=\big(\sum\limits_{i=1,i\neq k}^{K}\sum\limits_{n=1}^{T_i+U_i}l_{1,n}^{(k,i)}P_n^{(i)}+\sigma^2\big)\mathbf{I}_{T_k+U_k},
	\end{aligned}
	\end{equation}
	where $\mathbf{E}_{\mathcal{S}}$ is the expectation operator over small-scale fading $\mathcal{S} =\{\mathbf{S}_m^{(k,i)}|k,i =1, ..., K\}$, $\mathbf{P}^{(i)}$ is the power matrix of cluster $i$, i.e, $\mathbf{P}^{(i)}=\text{diag}\big\{P_{1}^{(i)},..., P_{T_i+U_i}^{(i)}\big\}$, which includes the power of user 1 and 2, i.e., $\mathbf{P}^{(i)}=\mathbf{P}_1^{(i)}+\mathbf{P}_2^{(i)}$ with $\mathbf{P}_m^{(i)}=\text{diag}\big\{P_{m,1}^{(i)},..., P_{m,T_i+U_i}^{(i)}\big\}, \ m=1,2$. Correspondingly, we denote $\sigma_{k,1}^2(\mathbf{P})$ as
	\begin{equation}
	\label{eq1}
	\sigma_{k,1}^2(\mathbf{P})=\sum\limits_{i=1,i \neq k}^{K}\sum\limits_{n=1}^{T_i+U_i}l_{1,n}^{(k,i)}P_{n}^{(i)}+\sigma^2,
	\end{equation}
	which is the total noise suffered by user 1 and $\mathbf{P}=\{\mathbf{P}^{(i)}|i=1,...,K\}$. 

	As for user 2, it regards user 1's signal as the background noise and directly decodes its own message. With the inter-cluster and inter-user interference, the interference plus noise of user 2 is calculated by
	\begin{equation}
	\sigma_{k,2}^2(\mathbf{P})=\sum\limits_{i=1,i \neq k}^{K}\sum\limits_{n=1}^{T_i+U_i}l_{2,n}^{(k,i)}P_{n}^{(i)}+\sum\limits_{n=1}^{T_k+U_k}l_{2,n}^{(k,k)}P_{1,n}^{(k)}+\sigma^2.
	\end{equation}
	When only large-scale CSI is known at transmitters, the ergodic sum rate is of great interest to this network
	\begin{equation}
			\bar{R}(\mathbf{P})=\sum\limits_{k=1}^{K}\sum\limits_{m=1}^{2}\bar{R}_{k,m}(\mathbf{P}), \\
	\end{equation}
	where 
	\begin{equation}
	\bar{R}_{k,m}(\mathbf{P})=\mathbf{E}_{\mathcal{S}}\bigg[\log_2\det\big(\mathbf{I}_{M_k}+\frac{\mathbf{H}_m^{(k,k)}\mathbf{P}_m^{(k)}\mathbf{H}_m^{(k,k)^H}}{\sigma_{k,m}^2(\mathbf{P})}\big)\bigg]
	\end{equation}
    is the ergodic rate of user $m$ in cluster $k$.
    
	To ensure the NOMA principle is applicable, user 1 should be capable to decode the message of user 2. In other words,
	the capacity that user 1 decodes the message of user 2, i.e., $\hat{\bar{R}}_{k,1}(\mathbf{P})$ should be greater than user 2's own capacity. This condition could be expressed as
	\begin{equation}
		\label{eqn28}
		\bar{R}_{k,2}(\mathbf{P})\leqslant \hat{\bar{R}}_{k,1}(\mathbf{P}), \ k=1,...,K,
	\end{equation} 
	where $\hat{\bar{R}}_{k,1}(\mathbf{P})$ is denoted as
	\begin{equation}
		\hat{\bar{R}}_{k,1}(\mathbf{P})= \mathbf{E}_{\mathbf{S}}\bigg[\log_2\det\big(\mathbf{I}_{M_k}+\frac{\mathbf{H}_1^{(k,k)}\mathbf{P}_2^{(k)}\mathbf{H}_1^{(k,k)^H}}{\hat{\sigma}_{k,1}^{2}(\mathbf{P})}\big)\bigg],
	\end{equation}
	\begin{equation}
	\hat{\sigma}_{k,1}^{2}(\mathbf{P})=\sum\limits_{i=1,i \neq k}^{K}\sum\limits_{n=1}^{T_i+U_i}l_{1,n}^{(k,i)}P_{n}^{(i)}+\sum\limits_{n=1}^{T_k+U_k}l_{1,n}^{(k,k)}P_{1,n}^{(k)}+\sigma^2.
\end{equation}
As we can see, the rate condition does not require user 1 dominates in all channels. Since the TBSs and UAVs are distributed in different positions, there is a complicated relationship between the decoding order and user positions. It is an interesting problem to investigate the pairing scheme and considerable gains can be achieved if we take both user pairing and resource allocation into account \cite{ad9}. For simplicity, in this paper, we focus on the case that user 1 and user 2 are separated by a large distance so that the constraint \eqref{eqn28} can be guaranteed. This consideration also accords with the wide sparsity of maritime IoT.

Since we assume user 1 possesses better channels,
traditional opportunistic schemes tend to give more resources to user 1 to exploit more efficiency. This greatly influences the fairness of the network. So, to meet the basic QoS requirement of user 2, i.e., $r_k$, we impose the following constraint
	\begin{equation}
	\label{eqn38}
	\bar{R}_{k,2}(\mathbf{P})\geqslant r_k, \ k=1,...,K.
	\end{equation}

	Furthermore, satellite users may suffer from inter-segment interference. To avoid potential impairment, the leakage interference is restricted not to exceed the maximum threshold of satellite receivers, i.e., $\bar{P}$,
	\begin{equation}
	\label{eq2}
	\sum\limits_{k=1}^K\mathbf{E}_{\mathbf{s'}}\parallel\mathbf{h'}^{(j,k)}\mathbf{P}^{(k)}\mathbf{h'}^{(j,k)^H}\parallel \leqslant \bar{P},  j=1,..., J,
	\end{equation}
	where $\mathbf{s'} =\{\mathbf{s'}^{(j,k)}|j=1,...,J, k=1, ..., K\}$. Based on the Gaussian distribution of $\mathbf{s'}$, the left side of \eqref{eq2} can be calculated as follows
	\begin{equation}
	\sum\limits_{k=1}^K\mathbf{E}_{\mathbf{s'}}\parallel\mathbf{h'}^{(j,k)}\mathbf{P}^{(k)}\mathbf{h'}^{(j,k)^H}\parallel=\sum\limits_{k=1}^K\sum_{n=1}^{T_k+U_k}{l'}_{n}^{(j,k)}P_{n}^{(k)}.
	\end{equation}
	Generally, the nearshore network also suffers from the interference from the satellite, but this impact is much smaller compared with strong inter-cluster and inter-user interference and we neglect it in this paper.
	
	With the consideration of the QoS requirement \eqref{eqn38}, inter-segment interference \eqref{eq2} and practical power constraints, we are aimed to maximize the ergodic sum rate of the nearshore network based on the large-scale CSI. 
	\begin{subequations}
	\label{P1}
	\begin{align}
		\mbox{(P1)} \ \max\limits_{\mathbf{P}}\
		&\bar{R}(\mathbf{P}) \label{eqd6}  \\
		s.t. \ &\bar{R}_{k,2}(\mathbf{P}) \geqslant r_k \label{eq4}\\
		&\mathbf{P}^{(k)}\succeq 0  \\
		&\operatorname{tr}(\mathbf{P}^{(k)})\leqslant P_{\max} \\
		&P_{n}^{(k)} \leqslant P_{\text{Tmax}},     n=1,...,T_k \\
		&P_{n}^{(k)} \leqslant P_{\text{Umax}},   n=T_k+1,...,T_k+U_k \\
		&\sum\limits_{k=1}^K\sum_{n=1}^{T_k+U_k}P_{n}^{(k)}{l'}_{n}^{(j,k)} \leqslant\bar{P},\ j=1,...,J \\
		&k=1,...,K, \nonumber
	\end{align}
	\end{subequations}
where  $P_{\text{Tmax}}$ and $P_{\text{Umax}}$ are the maximum affordable transmit power of TBSs and UAVs,  $P_{\max}$ is the power budget of each cluster. The problem is complex due to the non-convex objection and the non-convex feasible set. The ergodic rate is trapped in expectation operations, leading to a no closed-form expression. To tackle this problem, the tricky expectations should be first removed. By leveraging the random matrix theory,   $\bar{R}_{k,m}(\mathbf{P})$  can be approximated into a closed-form expression without the expectation operator \cite{ref3}
\begin{subequations} 
	\begin{equation}
		\!\!\!\!\!\!\!\!\!\!\!\!\!\!\!\!\!\!\!\!\!\!\!\!\!\!\!\!\!\!\!\!\!\!\!\!\!\!\!\!\!\!\!\!\!\!\!\!\!\!\bar{R}_{k,m}(\mathbf{P}) \approx g_{k,m}(\mathbf{P},v^*_{k,m}) \label{eqn1}
	\end{equation}	
	\begin{equation}
		\label{eq20}
		\begin{aligned}
			&=\sum\limits_{n=1}^{T_k+U_k}\log_2\big(1+\frac{M_kl_{m,n}^{(k,k)}P_{m,n}^{(k)}}
			{v^*_{k,m}\sigma_{k,m}^2(\mathbf{P})}\big) \\
		&	+M_k\log_2{v^*_{k,m}}-M_k\log_2e[1-(v^*_{k,m})^{-1}],
		\end{aligned}
	\end{equation}
\end{subequations} 
where $v^*_{k,m}$ is given by 
\begin{equation}
	\label{eqnn1}
	v^*_{k,m}=1+\sum_{n=1}^{T_k+U_k}\frac{l_{m,n}^{(k,k)}P_{m,n}^{(k)}}{\sigma^2_{k,m}(\mathbf{P})+M_kl_{m,n}^{(k,k)}P_{m,n}^{(k)}(v^*_{k,m})^{-1}}.
\end{equation}
This approximation is derived in the asymptotic case, e.g., the number of antennas goes to infinity. However, it has been shown quite accurate even with two or three antennas \cite{ref3}.

Then, the original problem (P1) is recast into (P2) as follows
\begin{subequations}
	\label{P2}
	\begin{align}
		\mbox{(P2)} \ \max\limits_{\mathbf{P}}\
		&g(\mathbf{P},\mathbf{v}^*)  \label{eqd7} \\
		s.t.\
		&g_{k,2}(\mathbf{P},v^*_{k,2}) \geqslant r_k \label{eqn24}\\
		&v^*_{k,m}=\label{eqd1}\\
		&1+\sum_{n=1}^{T_k+U_k}\frac{l_{m,n}^{(k,k)}P_{m,n}^{(k)}}{\sigma^2_{k,m}(\mathbf{P})+M_kl_{m,n}^{(k,k)}P_{m,n}^{(k)}(v^*_{k,m})^{-1}}  \nonumber\\
		&\mathbf{P}^{(k)}\succeq0  \\
		&\operatorname{tr}(\mathbf{P}^{(k)})\leqslant P_{\max} \\
		&P_{n}^{(k)} \leqslant P_{\text{Tmax}},     n=1,...,T_k \\
		&P_{n}^{(k)} \leqslant P_{\text{Umax}},   n=T_k+1,...,T_k+U_k \\
		&\sum\limits_{k=1}^K\sum_{n=1}^{T_k+U_k}P_{n}^{(k)}{l'}_{n}^{(j,k)} \leqslant\bar{P},\ j=1,...,J \\
		&m=1,2 \ \mbox{and}  \ k=1,...,K,\nonumber  
	\end{align}
\end{subequations}
where $g(\mathbf{P},\mathbf{v}^*)=\sum\limits_{k=1}^K\sum\limits_{m=1}^2g_{k,m}(\mathbf{P},v^*_{k,m})$ and $\mathbf{v}^*=\{v^*_{k,m}| m=1,2, \ k=1,...,K\}$. 	
Although we get rid of the expectation operations in (P2), the introduced implicit parameter  $\mathbf{v}^*$  with an extra group of complex non-linear constraints \eqref{eqd1} brings new challenges. The objection  \eqref{eqd7} and QoS constraint \eqref{eqn24} are both coupled with $\mathbf{v}^*$ in a non-convex form, which makes this problem  hard to tackle. To eliminate these non-linear constraints, we apply two different schemes for the objection and the QoS constraint, respectively.

\subsection{An Equivalent Max-Min Problem}
By analyzing the monotonicity of  $g_{k,m}(\mathbf{P},v_{k,m})$, it is proved that $ g_{k,m}(\mathbf{P},v_{k,m}) $ is monotonically decreasing in $[1, v^*_{k,m})$ and monotonically increasing in $[v^*_{k,m},+\infty)$ \cite{ref3}. It means $v^*_{k,m}$ is the minimal point of $g_{k,m}(\mathbf{P},v_{k,m})$, so, it can be expressed in an optimization form as
\begin{equation}
	g_{k,m}(\mathbf{P},v^*_{k,m})=\min_{v_{k,m}\geqslant 1}g_{k,m}(\mathbf{P},v_{k,m}). \label{eqd8}
\end{equation}
Based on \eqref{eqd8}, we thus recast (P2) into a max-min problem.
\begin{subequations}
	\label{P3}
	\begin{align}
		\mbox{(P3)} \max\limits_{\mathbf{P}}&\min\limits_{\mathbf{v}}\
		g(\mathbf{P},\mathbf{v})   \\
		s.t.\
		&g_{k,2}(\mathbf{P},v^*_{k,2}) \geqslant r_k \label{eqd4}\\
		& v^*_{k,2}=\label{eqd3}\\
		& \  1+\sum_{n=1}^{T_k+U_k}\frac{l_{2,n}^{(k,k)}P_{2,n}^{(k)}}{\sigma^2_{k,2}(\mathbf{P})+M_kl_{2,n}^{(k,k)}P_{2,n}^{(k)}(v^*_{k,2})^{-1}} \nonumber \\
		&v_{k,m} \geqslant 1 \\
		&\mathbf{P}^{(k)}\succeq0  \\
		&\operatorname{tr}(\mathbf{P}^{(k)})\leqslant P_{\max} \\
		&P_{n}^{(k)} \leqslant P_{\text{Tmax}},     n=1,...,T_k \\
		&P_{n}^{(k)} \leqslant P_{\text{Umax}},   n=T_k+1,...,T_k+U_k \\
		&\sum\limits_{k=1}^K\sum_{n=1}^{T_k+U_k}P_{n}^{(k)}{l'}_{n}^{(j,k)} \leqslant\bar{P},\ j=1,...,J \\
		&m=1,2 \ \mbox{and}  \ k=1,...,K.\nonumber 
	\end{align}
\end{subequations}
It is easy to find the optimal solution $(\mathbf{P}^*,\mathbf{v}^*)$ is the saddle point of $g(\mathbf{P},\mathbf{v})$. Intuitively, the QoS constraint \eqref{eqd4} can also be transformed as
\begin{equation}
	\min\limits_{v_{k,2}\geqslant 1} g_{k,2}(\mathbf{P},v_{k,2}) \geqslant r_k,
\end{equation}
and further equivalently turn into
\begin{equation}
	g_{k,2}(\mathbf{P},v_{k,2}) \geqslant r_k \label{eqd2}.
\end{equation}
However, this transformation actually extends the optimization space while making no sense for further simplification. So, we turn to another way to tackle the QoS constraint. By using the method of auxiliary function, a critical monotonically function is introduced, so as to further simplify this problem.
%

\subsection{Simplification of the Non-Linear Constraint} 	
Define	$f_{k}(\mathbf{P},z_k)$ as 
\begin{equation}
	\begin{aligned}
		f_{k}(\mathbf{P},z_k)
		=\sum\limits_{n=1}^{T_k+U_k}\bigg(\log_2\big(1+\frac{M_kl_{2,n}^{(k,k)}P_{2,n}^{(k)}}{z_k\sigma^2_{k,2}(\mathbf{P})}\big)&\\				-\frac{\log_2(e)M_kl_{2,n}^{(k,k)}P_{2,n}^{(k)}}{z_k\sigma^2_{k,2}(\mathbf{P})+M_kl_{2,n}^{(k,k)}P_{2,n}^{(k)}}\bigg)+M_k\log_2z_k&.
	\end{aligned}
\end{equation}

\begin{theorem}
	\label{1}
	Under the condition that $ M_k\geqslant T_k+U_k$, $f_{k}(\mathbf{P},z_k)$ is monotonically increasing  with respect to $z_k$. The relationship between $g_{k,2}(\mathbf{P},v^*_{k,2})$ and $f_{k}(\mathbf{P},z_k)$ is given by
	\begin{equation}
		\label{eq30}
		g_{k,2}(\mathbf{P},v^*_{k,2})=\max\limits_{1\leqslant z_k\leqslant v^*_{k,2}}f_{k}(\mathbf{P},z_k).
	\end{equation}	
	
	\textit{Proof}:\  See Appendix A.    
\end{theorem}
According to \eqref{eq30} and the monotonicity of $f_{k}(\mathbf{P},z_k)$, the QoS constraint \eqref{eqd4} can be simplified as
\begin{equation}
	\label{eqd5}
	\begin{aligned}
		\max\limits_{1 \leqslant z_k\leqslant v^*_{k,2}}f_{k}(\mathbf{P},z_k) \geqslant r_k \Rightarrow
		\left\{
		\begin{array}{l}
			f_{k}(\mathbf{P},z_k) \geqslant r_k, \\
			1 \leqslant z_k\leqslant v^*_{k,2}.
		\end{array}
	\right.
	\end{aligned}	
\end{equation}
Note that  \eqref{eqd3} can be equivalently rewritten as 
\begin{equation}
	(v^*_{k,2})^{-1}+\sum_{n=1}^{T_k+U_k}\frac{l_{2,n}^{(k,k)}P_{2,n}^{(k)}}{v^*_{k,2}\sigma^2_{k,2}(\mathbf{P})+M_kl_{2,n}^{(k,k)}P_{2,n}^{(k)}}=1.
\end{equation}
As $(z_{k})^{-1}+\sum\limits_{n=1}^{T_k+U_k}\frac{l_{2,n}^{(k,k)}P_{2,n}^{(k)}}{z_{k}\sigma^2_{k,2}(\mathbf{P})+M_kl_{2,n}^{(k,k)}P_{2,n}^{(k)}}$ is  monotonically
decreasing with $z_{k}$, so, for any $z_k\geqslant 0$, we have
\begin{equation}
	\label{eqn31}
	\begin{aligned}
		z_k &\leqslant v^*_{k,2} \Rightarrow \\ &(z_k)^{-1}+\sum_{n=1}^{T_k+U_k}\frac{l_{2,n}^{(k,k)}P_{2,n}^{(k)}}{z_k\sigma^2_{k,2}(\mathbf{P})+M_kl_{2,n}^{(k,k)}P_{2,n}^{(k)}} \geqslant 1. \\
	\end{aligned}
\end{equation}
Based on \eqref{eqd5} and \eqref{eqn31}, (P3) is further recast into (P4).

\begin{subequations}
	\label{P4}
	\begin{align}
		\mbox{(P4)} \ \max\limits_{\mathbf{P},z_k}&\min\limits_{\mathbf{v}}\
		g(\mathbf{P},\mathbf{v})   \\
		s.t.\
		&f_{k}(\mathbf{P},z_k)\geqslant r_k \label{eqn8}\\
		&(z_k)^{-1}+ \\
		&\ \ \ \ \ \sum_{n=1}^{T_k+U_k}\frac{l_{2,n}^{(k,k)}P_{2,n}^{(k)}}{z_k\sigma^2_{k,2}(\mathbf{P})+M_kl_{2,n}^{(k,k)}P_{2,n}^{(k)}} \geqslant 1 \nonumber\\ 
		&v_{k,m} \geqslant 1, z_k \geqslant 1\label{eqn7}\\
		&\mathbf{P}^{(k)}\succeq0  \\
		&\operatorname{tr}(\mathbf{P}^{(k)})\leqslant P_{\max} \\
		&P_{n}^{(k)} \leqslant P_{\text{Tmax}},     n=1,...,T_k \\
		&P_{n}^{(k)} \leqslant P_{\text{Umax}},   n=T_k+1,...,T_k+U_k \\
		&\sum\limits_{k=1}^K\sum_{n=1}^{T_k+U_k}P_{n}^{(k)}{l'}_{n}^{(j,k)} \leqslant\bar{P},\ j=1,...,J\\
		&m=1,2 \ \mbox{and}  \ k=1,...,K.\nonumber 
	\end{align}
\end{subequations}
Take a further look of $f_{k}(\mathbf{P},z_k)$, it can be reorganized as
\begin{equation}
	\label{eqn30}
	\begin{aligned}
		&f_{k}(\mathbf{P},z_k)
		=-\log_2(e)\sum\limits_{n=1}^{T_k+U_k}\bigg(\frac{M_kl_{2,n}^{(k,k)}P_{2,n}^{(k)}}{z_k\sigma^2_{k,2}(\mathbf{P})+M_kl_{2,n}^{(k,k)}P_{2,n}^{(k)}}+\\
		&\text{In}\big(1-\frac{M_kl_{2,n}^{(k,k)}P_{2,n}^{(k)}}{z_k\sigma^2_{k,2}(\mathbf{P})+M_kl_{2,n}^{(k,k)}P_{2,n}^{(k)}}\big)\bigg)
		+M_k\log_2z_k,
	\end{aligned}
\end{equation}
where the fraction term $\frac{M_kl_{2,n}^{(k,k)}P_{2,n}^{(k)}}{z_k\sigma^2_{k,2}(\mathbf{P})+M_kl_{2,n}^{(k,k)}P_{2,n}^{(k)}}$ happens to be the same. Motivated by this, we introduce a group of slack variables $\mathbf{t}_k=\{t_{k,n} |n=1,...,T_k+U_k\}$,  and define $\mathcal{F}_k(z_k,\mathbf{t}_k)$ as follows
\begin{equation}
	\label{ad8}
	\begin{split}
		\mathcal{F}_k(z_k,\mathbf{t}_k)=&-\log_2(e)\sum\limits_{n=1}^{T_k+U_k}\big(t_{k,n}+\mbox{In}\big(1-t_{k,n})\big) \\
		&+M_k\log_2z_k, \ k=1,...,K.
	\end{split}
\end{equation}
Based on the increasing monotonicity of $-[x+\text{In}(1-x)]$, it is clear that $\mathcal{F}_k(z_k,\mathbf{t}_k)$ is
monotonically increasing with $\mathbf{t}_k$ and thus problem (P4) is turned into (P5)
\begin{subequations}
	\label{P5}
	\begin{align}
		\mbox{(P5)} \max\limits_{\mathbf{P}, z_k,\mathbf{t}_k}&\min\limits_{\mathbf{v}}\
		g(\mathbf{P},\mathbf{v})   \label{eqn26}\\
		s.t.\
		&\mathcal{F}_k(z_k,\mathbf{t}_k)\geqslant r_k \label{eqn25}\\
		&(z_k)^{-1}+\frac{1}{M_k}\sum_{n=1}^{T_k+U_k}t_{k,n} \geqslant 1  \label{ad3}\\
		& t_{k,n}\leqslant \frac{M_kl_{2,n}^{(k,k)}P_{2,n}^{(k)}}{z_k\sigma^2_{k,2}(\mathbf{P})+M_kl_{2,n}^{(k,k)}P_{2,n}^{(k)}} \label{ad4} \\
		&v_{k,m} \geqslant 1, z_k\geqslant 1, t_{k,n}\geqslant 0  \label{ad10} \\
		&\mathbf{P}^{(k)}\succeq0  \\
		&\operatorname{tr}(\mathbf{P}^{(k)})\leqslant P_{\max} \\
		&P_{n}^{(k)} \leqslant P_{\text{Tmax}},     n=1,...,T_k \\
		&P_{n}^{(k)} \leqslant P_{\text{Umax}},   n=T_k+1,...,T_k+U_k \\
		&\sum\limits_{k=1}^K\sum_{n=1}^{T_k+U_k}P_{n}^{(k)}{l'}_{n}^{(j,k)}\leqslant\bar{P}, \ j=1,...,J \\
		&m=1,2, \ n=1,...,T_k+U_k, \ k=1,...,K.\nonumber 
	\end{align}
\end{subequations}
It is noted that if the constraint \eqref{eqn24} in (P2) is tight, the optimal solution of (P5) is obtained if and only if  $\mathcal{F}_k({z_k},\mathbf{t}_k)$ is maximized, which indicates that the equality holds in \eqref{eqn25}, \eqref{ad3} and \eqref{ad4}, and thus  $\mathcal{F}_k({z_k},\mathbf{t}_k)=g_{k,2}(\mathbf{P},v_{k,2}^*)$. So, we could draw the following lemma.
\begin{lemma}
	The optimal solution to (P5) is also the optimal solution to (P2) and vice versa.

	\textit{Proof}:\  See Appendix B.    
\end{lemma}

Observing (P5), the objection function is still non-convex with respect to $\mathbf{P}$ and $\mathbf{v}$  and the constraint \eqref{ad4} is shown in a complex fraction form. Proceeding with these problems, we firstly introduce a group of variables $\mathbf{w}=\{w_{k,m}| m=1,2, \ k=1,...,K\}$, which is given by $e^{w_{k,m}}=v_{k,m}$. The quasi-convex function $g(\mathbf{P},\mathbf{v})$ with respect to 
$\mathbf{v}$ is then transformed into a convex function $\mathcal{G}(\mathbf{P},\mathbf{w})$ with respect to $\mathbf{w}$ \cite{ref3}, which is given by 
\begin{equation}
	\label{eqd16}
	\mathcal{G}(\mathbf{P},\mathbf{w})=\sum\limits_{k=1}^K\sum\limits_{m=1}^2\mathcal{G}_{k,m}(\mathbf{P},{w_{k,m}}),
\end{equation}
where
\begin{equation}
	\begin{aligned}
		\mathcal{G}_{k,m}(\mathbf{P},{w_{k,m}})&=\sum\limits_{n=1}^{T_k+U_k}\log_2\bigg(1+\frac{M_kl_{m,n}^{(k,k)}P_{m,n}^{(k)}}
		{e^{{w_{k,m}}}\sigma_{k,m}^2(\mathbf{P})}\bigg) \\
		&+M_k\log_2(e)[{w_{k,m}}+e^{-{w_{k,m}}}-1].
	\end{aligned}
\end{equation}	
In addition, $\mathcal{G}_{k,m}(\mathbf{P},{w_{k,m}})$ can be reorganized into the subtraction of two parts
\begin{equation}
	\label{eqn33}
	\mathcal{G}_{k,m}(\mathbf{P},{w_{k,m}})=\mathcal{G}_{k,m,1}(\mathbf{P},{w_{k,m}})-\mathcal{G}_{k,m,2}(\mathbf{P}),
\end{equation}
where
\begin{subequations}
	\begin{equation}
		\begin{aligned}
			\mathcal{G}_{k,m,1}(\mathbf{P},{w_{k,m}})=&\sum\limits_{n=1}^{T_k+U_k}\log_2\bigg(\sigma_{k,m}^2(\mathbf{P})+\frac{M_kl_{m,n}^{(k,k)}P_{m,n}^{(k)}}{e^{{w_{k,m}}}}\bigg)\\
			&+M_k\log_2e[{w_{k,m}}+e^{-{w_{k,m}}}-1],
		\end{aligned}
	\end{equation}
	\begin{equation}
		\mathcal{G}_{k,m,2}(\mathbf{P})=(T_k+U_k)\log_2\big(\sigma_{k,m}^2(\mathbf{P})\big),
	\end{equation}
\end{subequations}
and they are both concave with respect to $\mathbf{P}$. To decouple different variables in \eqref{ad4}, the logarithmic operation is applied. Then, this constraint is  reorganized as
\begin{equation}
	\label{eqn27}
	\begin{aligned}
		\mbox{In}\big(M_kP_{2,n}^{(k)}l_{2,n}^{(k,k)}\big)+&\mbox{In}(1-t_{k,n})\geqslant \\
		&\mbox{In}(z_k)+\mbox{In}(t_{k,n})+\mbox{In}({\sigma}^{2}_{k,2}(\mathbf{P})).
	\end{aligned}
\end{equation}
	Consequently, according to \eqref{eqd16} and \eqref{eqn27}, problem (P5) comes into (P6) 
	\begin{subequations}
	\label{P6}
	\begin{align}
		\mbox{(P6)}  \max\limits_{\mathbf{P}, z_k,\mathbf{t}_k}\min\limits_{\mathbf{w}}
		&\ \mathcal{G}(\mathbf{P},\mathbf{w}) \label{eqn20} \\
		s.t.\
		&\mathcal{F}_k(z_k,\mathbf{t}_k)\geqslant r_k\label{eqn34}\\
		&(z_k)^{-1}+\frac{1}{M_k}\sum_{n=1}^{T_k+U_k}t_{k,n} \geqslant 1  \label{eqn29}\\
		&\mbox{In}\big(M_kP_{2,n}^{(k)}l_{2,n}^{(k,k)}\big)+\mbox{In}(1-t_{k,n})\geqslant \nonumber\\
		&\mbox{In}(z_k)+\mbox{In}(t_{k,n})+\mbox{In}({\sigma}^{2}_{k,2}(\mathbf{P})) \label{eqn19}\\
		&w_{k,m}\geqslant 0, z_k\geqslant 1, t_{k,n}\geqslant 0\\
		&\mathbf{P}^{(k)}\succeq0  \\
		&\operatorname{tr}(\mathbf{P}^{(k)})\leqslant P_{\max} \\
		&P_{n}^{(k)} \leqslant P_{\text{Tmax}},     n=1,...,T_k\\
		&P_{n}^{(k)} \leqslant P_{\text{Umax}},   n=T_k+1,...,T_k+U_k\\
		&\sum\limits_{k=1}^K\sum_{n=1}^{T_k+U_k}P_{n}^{(k)}{l'}_{n}^{(j,k)} \leqslant\bar{P}, j=1,...,J \\
		&m=1,2, \ n=1,...,T_k+U_k, \ k=1,...,K.\nonumber 
	\end{align}
	\end{subequations}
	Up to now, all the separate terms in (P6) are shown in convex or concave forms. 
	By applying the method of successive convex approximation, $\mathcal{G}_{k,m,2}(\mathbf{P})$ in \eqref{eqn20}, $\log_2(1-t_{k,n})$ in  \eqref{eqn34}, $(z_k)^{-1}$ in \eqref{eqn29}, and $\mbox{In}(z_k)+\mbox{In}(t_{k,n})+\mbox{In}({\sigma}^{2}_{k,2}(\mathbf{P}))$ in \eqref{eqn19}
	should be linearized by their first-order Taylor expansions. Given points $\mathbf{P}^{s-1}$, $z_k^{s-1}$ and $\mathbf{t}_k^{s-1}$ of the last iteration, the convex subproblem of iteration $s$ can be formulated as 
	\begin{subequations}
	\label{P7}
	\begin{align}
		\mbox{(P7)}\ &\max\limits_{\mathbf{P},z_k,\mathbf{t}_k}\min\limits_{\mathbf{w} } \ \mathcal{G}(\mathbf{P},\mathbf{w}|\mathbf{P}^{s-1})   \\		
		s.t.  \ \
		&\mathcal{F}_k(z_k,\mathbf{t}_k|\mathbf{t}_k^{s-1}) \geqslant r_k \label{eq40}\\
		&-\frac{z_k}{(z_k^{s-1})^2}+\frac{2}{z_k^{s-1}}+\frac{1}{M_k}\sum_{n=1}^{T_k+U_k}t_{k,n}\geqslant 1 \label{eq41}\\
		&\mbox{In}\big(M_kP_{2,n}^{(k)}l_{2,n}^{(k,k)}\big)+\mbox{In}(1-t_{k,n})\geqslant \label{eq42}\\
		&\mbox{In}(z_k|z_k^{s-1})+\mbox{In}(t_{k,n}|t_{k,n}^{s-1})+\mbox{In}(\sigma^2_{k,2}(\mathbf{P}|\mathbf{P}^{s-1}))\nonumber\\		
		& w_{k,m}\geqslant 0, z_k\geqslant 1, t_{k,n}\geqslant 0\\
		&\mathbf{P}^{(k)}\succeq0  \\
		&\operatorname{tr}(\mathbf{P}^{(k)})\leqslant P_{\max}\\
		&P_{n}^{(k)} \leqslant P_{\text{Tmax}},     n=1,...,T_k\\
		&P_{n}^{(k)} \leqslant P_{\text{Umax}},   n=T_k+1,...,T_k+U_k\\
		&\sum\limits_{k=1}^K\sum_{n=1}^{T_k+U_k}P_{n}^{(k)}{l'}_{n}^{(j,k)} 	\leqslant\bar{P},  j=1,...,J\\
		&m=1,2, \ n=1,...,T_k+U_k, \ k=1,...,K,\nonumber 
	\end{align}
	\end{subequations}
	where the  expressions $\mathcal{G}(\mathbf{P},\mathbf{w}|\mathbf{P}^{s-1})$,
	$\mathcal{F}_k(z_k,\mathbf{t}_k|\mathbf{t}_k^{s-1})$ and $\mbox{In}(z_k|z_k^{s-1})+\mbox{In}(t_{k,n}|t_{k,n}^{s-1})+\mbox{In}(\sigma^2_{k,2}(\mathbf{P}|\mathbf{P}^{s-1}))$ as well as $\sigma_{k,1}^2(\mathbf{P}|\mathbf{P}^{s-1})$, $ {\sigma}_{k,2}^2(\mathbf{P}|\mathbf{P}^{s-1})$ are  shown at the top of the next page. The details of the algorithm are summarized in \textbf{Algorithm \ref{Tab1}}, the convergence of which is shown in the following \textit{Lemma 2}.

\begin{lemma}
	\label{2}
	\textbf{Algorithm \ref{Tab1}} is assured to converge, at least to a local optimum.\\
	
	\textit{Proof}:\  See Appendix C.    
\end{lemma}

	\begin{algorithm}[t]
	\caption{Iterative power allocation algorithm.}
	\label{Tab1}
	\begin{algorithmic}[1]
		\REQUIRE {$P_{\max}$, $P_{\text{Tmax}}$, $P_{\text{Umax}}$, $\bar{P}$, $r_k \ (k=1,...,K)$				
			\STATE \emph{Initialization}: set $\epsilon=1\times10^{-3}$}, $(P^{(k)})^0=P_{\max}$, \\ $(P_n^{(k)})^0=\frac{(P^{(k)})^0}{T_k+U_k},$ 
		$n=1,..., T_k+U_k\ \text{and} \ k=1,..,K.$ \\		
		\FOR {$j=1\sim J$}
		\WHILE { $(\sum\limits_{k=1}^K\sum\limits_{n=1}^{T_k+U_k}{l'}_{n}^{(j,k)}(P_{n}^{(k)})^0 > \bar{P})$}
		\STATE  $(P^{(k)})^0=(1-\epsilon)(P^{(k)})^0$;
		\STATE  $(P_n^{(k)})^0=\frac{(P^{(k)})^0}{T_k+U_k}$, $n=1, ..., T_k+U_k, k=1,..., K$;
		\ENDWHILE
		\ENDFOR	
		\FOR {$k=1\sim K, n=1\sim T_k$}
		\IF {$((P_n^{(k)})^0 > P_{\text{Tmax}})$}
		\STATE $(P_n^{(k)})^0=P_{\text{Tmax}}$, 
		\ENDIF
		\ENDFOR
		\FOR {$k=1\sim K, n=T_k+1\sim T_k+U_k$}
		\IF {$((P_n^{(k)})^0 > P_{\text{Umax}})$}
		\STATE $(P_n^{(k)})^0=P_{\text{Umax}}$;
		\ENDIF
		\ENDFOR
		\STATE $(P_{m,n}^{(k)})^0=\frac{(P_n^{(k)})^0}{2}, m=1,2,\ n=1,...,T_k+U_k, \text{and} \ k=1,...,K$,
		\STATE Calculate  $z_k^0,  \ \mathbf{t}_k^0, \ k=1,...,K$	based on \eqref{ad3} and \eqref{ad4} under the equality condition.
		\STATE \emph{Iterations:} $s=1, 2,...,$
		\REPEAT
		\STATE  Update  $\mathbf{P}^s$, $\mathbf{t}_k^s$ and $z_k^s, k=1,...,K$ by solving (P7);
		\UNTIL{$\frac{\left|\mathcal{G}(\mathbf{P}^s,\mathbf{w}^s)-\mathcal{G}(\mathbf{P}^{s-1},\mathbf{w}^{s-1})\right |}{\mathcal{G}(\mathbf{P}^{s-1},\mathbf{w}^{s-1})} \ \leqslant \epsilon$.}
		\ENSURE	 $\mathbf{P}^s.$ 
	\end{algorithmic}
\end{algorithm}

	\begin{figure*}[!t] 
	\normalsize 
	\begin{subequations}
		\begin{align}
			\sigma_{k,1}^2(\mathbf{P}|\mathbf{P}^{s-1})=&\sum\limits_{i=1,i \neq k}^{K}\sum\limits_{n=1}^{T_i+U_i}l_{1,n}^{(k,i)}\big(P_{n}^{(i)}-(P_{n}^{(i)})^{s-1}\big)+\sigma^2\\	
			\sigma_{k,2}^2(\mathbf{P}|\mathbf{P}^{s-1})=&\sum\limits_{i=1,i \neq k}^{K}\sum\limits_{n=1}^{T_i+U_i}l_{2,n}^{(k,i)}\big(P_{n}^{(i)}-(P_{n}^{(i)})^{s-1}\big)+\sum\limits_{n=1}^{T_k+U_k}l_{2,n}^{(k,k)}\big(P_{1,n}^{(k)}-(P_{1,n}^{(k)})^{s-1}\big)+\sigma^2
		\end{align}
	\end{subequations}

	\begin{subequations}
		\begin{equation}
			\mathcal{G}(\mathbf{P},\mathbf{w}|\mathbf{P}^{s-1})=\sum\limits_{k=1}^K\sum\limits_{m=1}^2\mathcal{G}_{k,m}(\mathbf{P},{w_{k,m}}|\mathbf{P}^{s-1})
		\end{equation}	
		\begin{equation}
			\begin{aligned}
				\mathcal{G}_{k,m}(\mathbf{P},{w_{k,m}}|\mathbf{P}^{s-1})&=\mathcal{G}_{k,m,1}(\mathbf{P},w_{k,m})-\mathcal{G}_{k,m,2}(\mathbf{P}|\mathbf{P}^{s-1})\\
				&=\mathcal{G}_{k,m,1}(\mathbf{P},w_{k,m})-\mathcal{G}_{k,m,2}(\mathbf{P}^{s-1})-(T_k+U_k)\log_2(e)\frac{\sigma^2_{k,m}(\mathbf{P}|\mathbf{P}^{s-1})}{\sigma^2_{k,m}(\mathbf{P}^{s-1})}
			\end{aligned}
		\end{equation}
	\end{subequations}

	\begin{equation}
		\begin{aligned}
			&\mathcal{F}_k(z_k,\mathbf{t}_k|\mathbf{t}_k^{s-1})=M_k\log_2z_k+\sum\limits_{n=1}^{T_k+U_k}\bigg(\log_2(e)\frac{t_{k,n}^{s-1}(t_{k,n}-1)}{1-t_{k,n}^{s-1}}-\log_2(1-t_{k,n}^{s-1})\bigg)
		\end{aligned}
	\end{equation}

	\begin{equation}
		\mbox{In}\big(z_k|z_k^{s-1}\big)+\mbox{In}\big(t_{k,n}|t_{k,n}^{s-1}\big)+\mbox{In}\big(\sigma^2_{k,2}(\mathbf{P}|\mathbf{P}^{s-1})\big)=\mbox{In}\big(z_k^{s-1}t_{k,n}^{s-1}\sigma^2_{k,2}(\mathbf{P}^{s-1})\big)+\frac{z_k}{z_k^{s-1}}+\frac{t_{k,n}}{t_{k,n}^{s-1}}+\frac{{\sigma}_{k,2}^2(\mathbf{P}|\mathbf{P}^{s-1})}{{\sigma}^2_{k,2}(\mathbf{P}^{s-1})}-2
	\end{equation}
	\vspace*{4pt} 
	\end{figure*}

	\begin{figure}
	\centering
	\includegraphics[width=3.0 in]{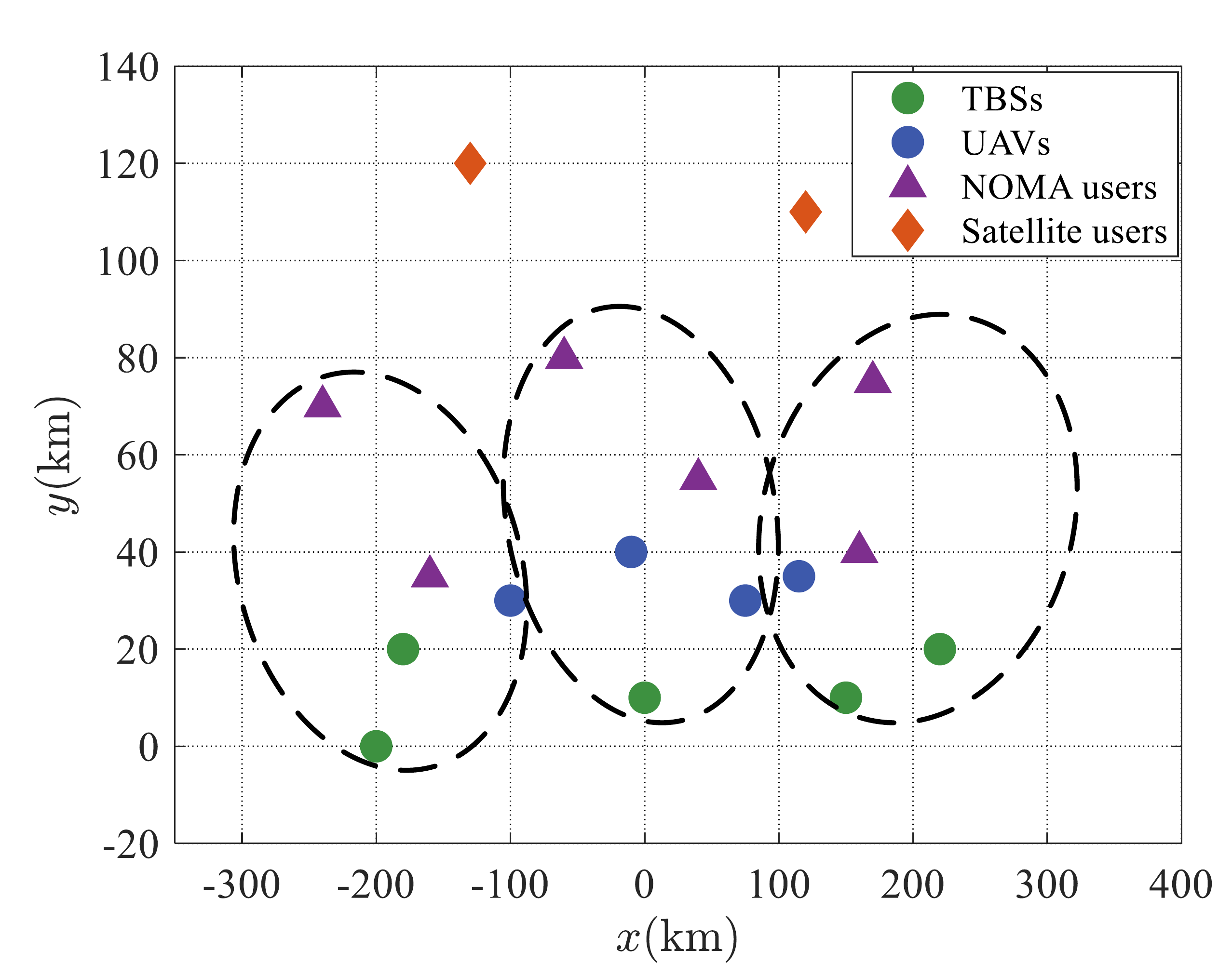}
	\caption{Illustration of the network topology for simulations.}
	\label{fig0}
	\end{figure}

	\begin{figure}
	\centering
	\includegraphics[width=3.3 in]{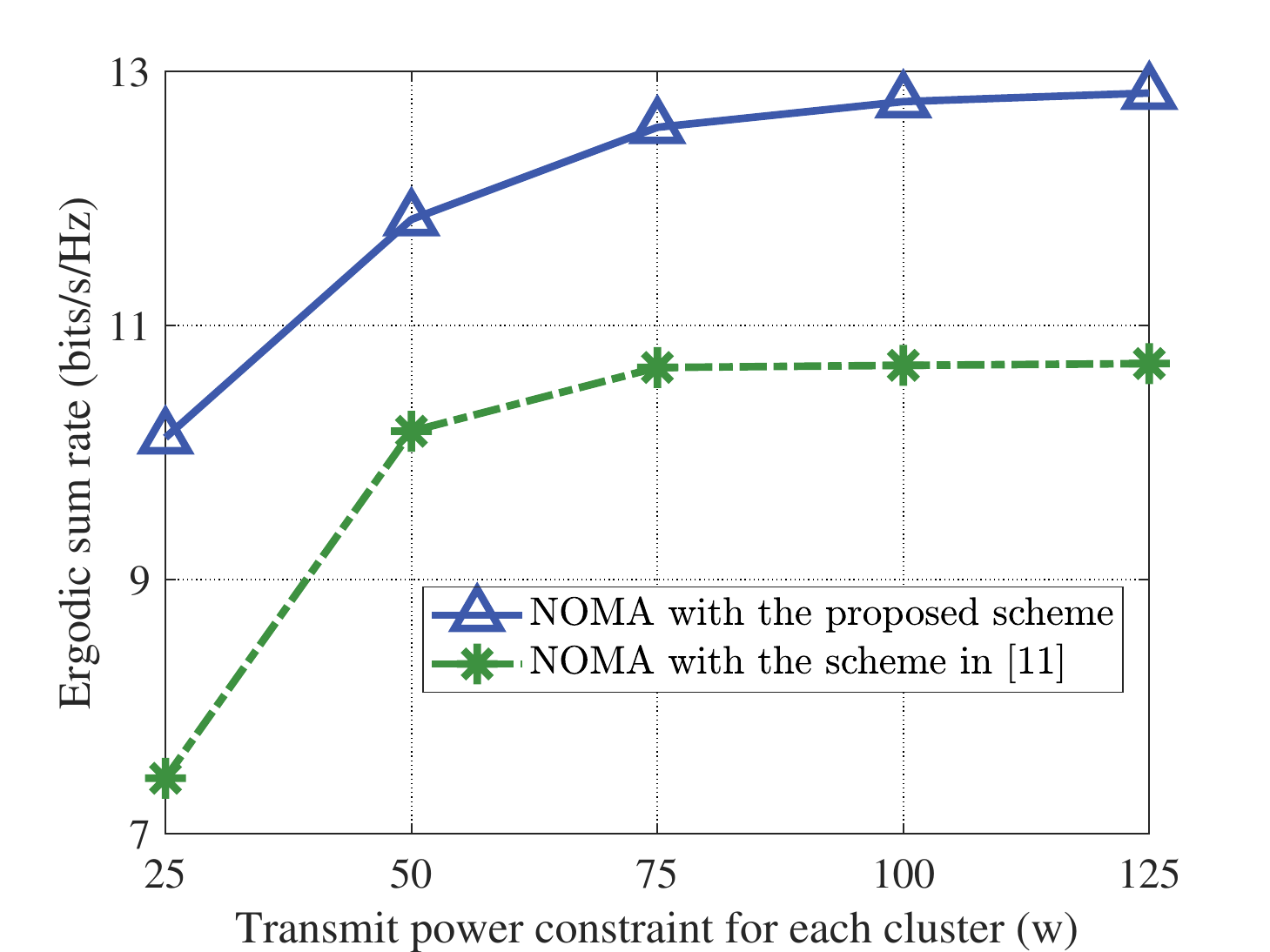}
	\caption{The ergodic sum rate with different power allocation schemes under the NOMA regime.}
	\label{fig1}
	\end{figure}

	\section{Simulation Results and Discussions}
	\label{section 4}
	
	In this section, we evaluate the proposed scheme through simulations. Without loss of generality, we assume there are three nearshore clusters, i.e., $K=3$, consisting of $T_1=T_3=2, \ T_2=1$ TBSs, $U_1=U_3=1, \ U_2=2$ UAVs. In each cluster, there are two NOMA users that are within 100km from the coastline. Both TBSs and UAVs are equipped with a single omnidirectional antenna, NOMA users are equipped with three antennas, i.e., $M_1=M_2=M_3=3$, and the antenna gains are given by $g_t=30\text{dBi}, \ g_u=10\text{dBi}$. The minimum rate requirement of user 2 is 0.5 bits/s/Hz, i.e., $r_1=r_2=r_3=0.5$ bits/s/Hz. Beyond clusters, two satellite users are also considered in our settings, i.e., $J=2$. The maximum tolerable interference of satellite users is $-100\text{dBm}$, i.e., $\bar{P}=-100\text{dBm}$, the maximum transmit power of TBSs/UAVs is $40$w and $30$w, i.e., $P_{\text{Tmax}}=40$w, $P_{\text{Umax}}=30$w and the maximum transmit power for each cluster changes from $25$w to $125$w. The heights of UAVs, TBSs and user terminals are assumed to be $h_u=800$m, $h_t=200$m and  $h_r=5$m, respectively. The constants are given by $c=3\times10^8$m/s, $\sigma^2=-107$dBm, $f_c=2$GHz, $a=5.0188$, $b=0.3511$, $c=3\times10^8$m/s, $\eta_{LOS}=0.1$, $\eta_{NLOS}=21$ and $C=1$.

	To make sure the NOMA scheme is applicable, we choose a topology where user 1 and user 2 are far apart so that the condition  \eqref{eqn28} is guaranteed. The specific distribution is shown in Fig. \ref{fig0} where the positive direction of $y$ points to the sea.

	Firstly, we evaluate the proposed scheme and compare it with the scheme in \cite{ref4}. As we can see from Fig. \ref{fig1}, the proposed scheme achieves much higher rates than that of \cite{ref4}. We can also find that when the maximum transmit power exceeds $75$w, increasing power cannot increase the sum rate under the scheme in \cite{ref4}. This means the system is saturated due to complex interference. For the proposed scheme, the saturated point does not yet occur in this figure. It can be explained that, based on the large-scale CSI, the system could flexibly allocate power to achieve a good balance between useful signals and harmful  interference.

	\begin{figure}
	\centering
	\includegraphics[width=3.3 in]{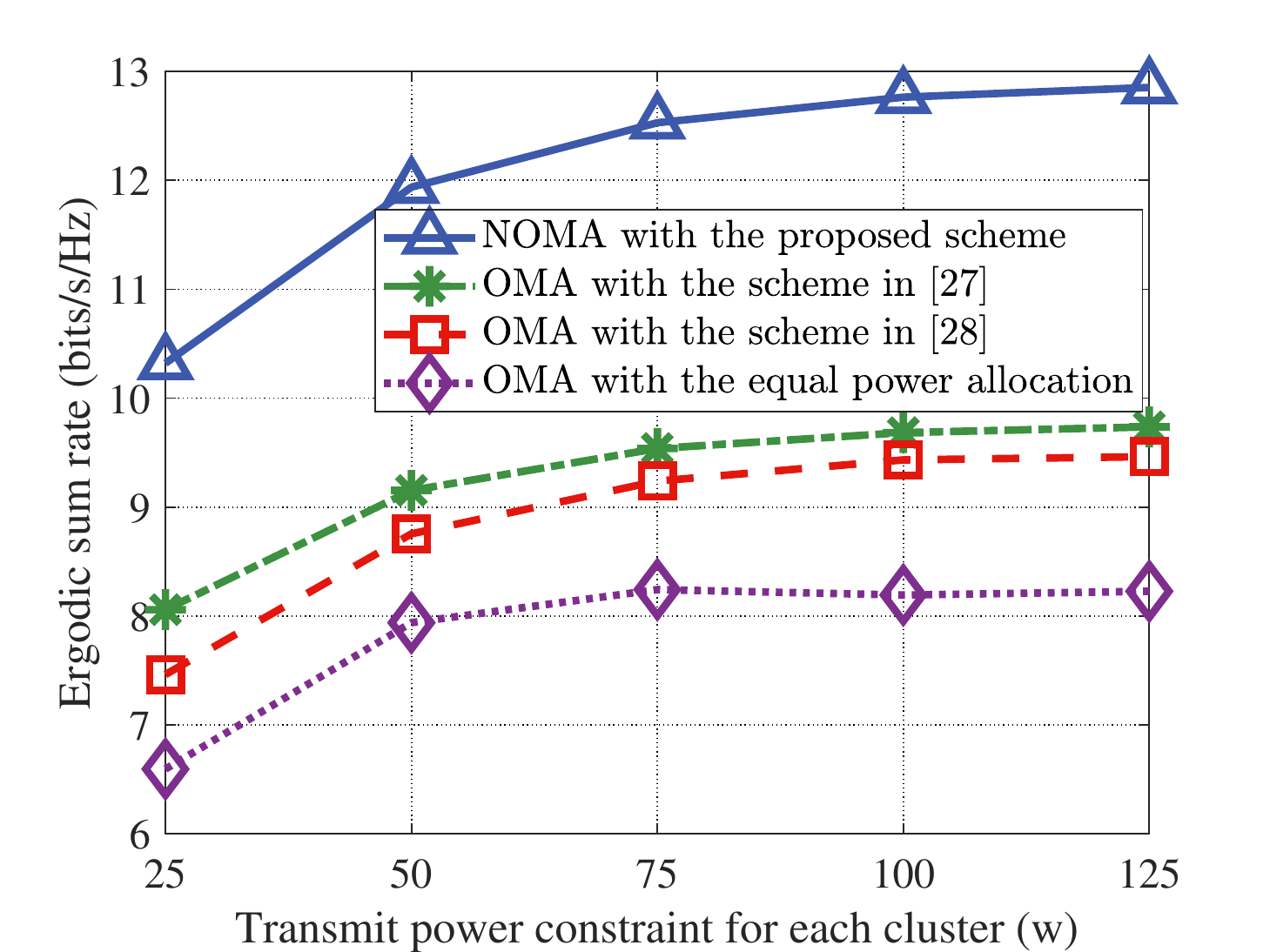}
	\caption{The ergodic sum rate with different power allocation schemes, under both NOMA and OMA regimes.}
	\label{fig2}
	\end{figure} 
	\begin{figure}
	\centering
	\includegraphics[width=3.3 in]{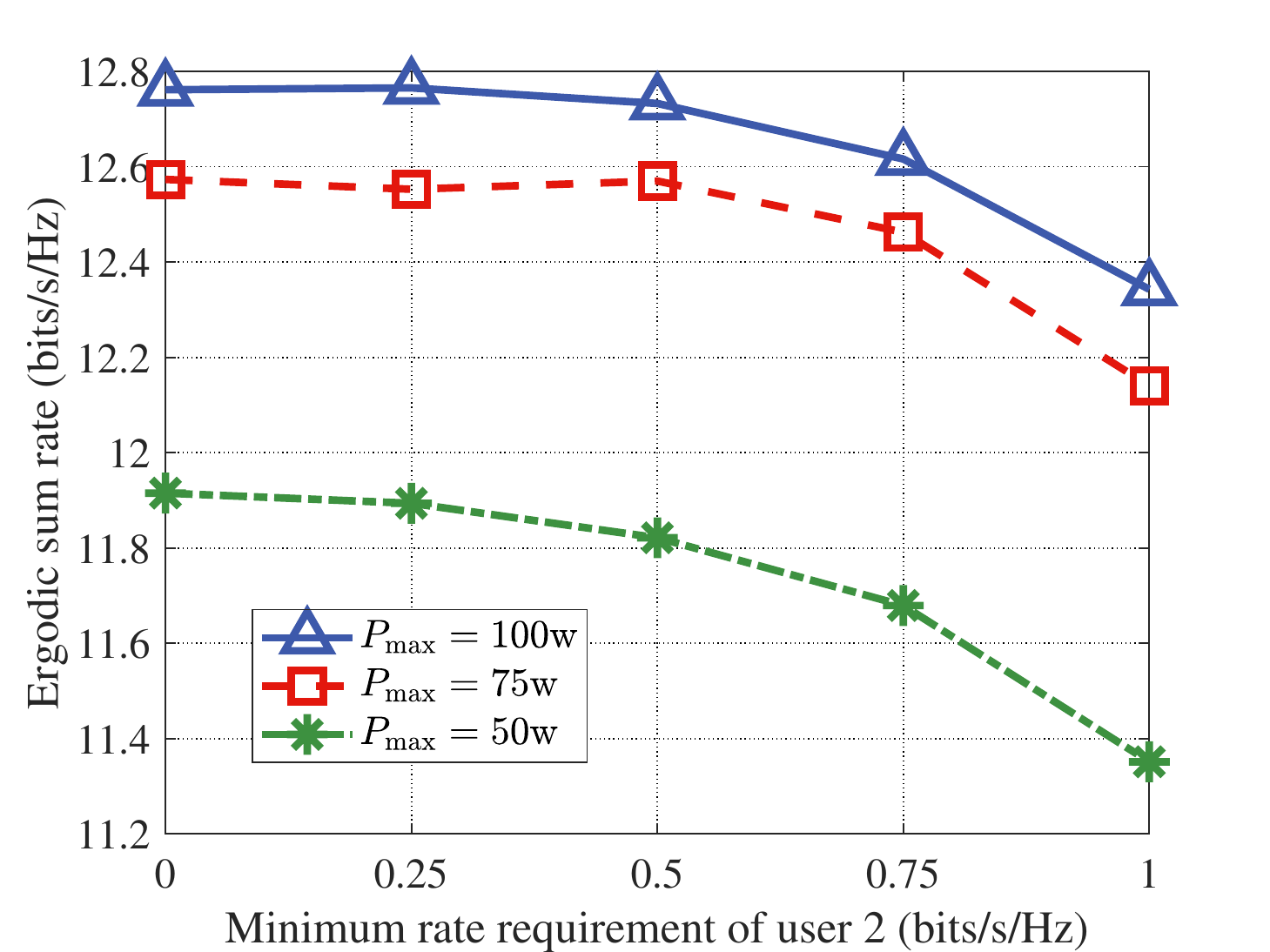}
	\caption{The ergodic sum rate with  different transmit power constraints, varying with the minimum rate requirement of user 2.}
	\label{fig4}
	\end{figure}

	\begin{figure}[t]
	\centering
	\includegraphics[width=3.3 in]{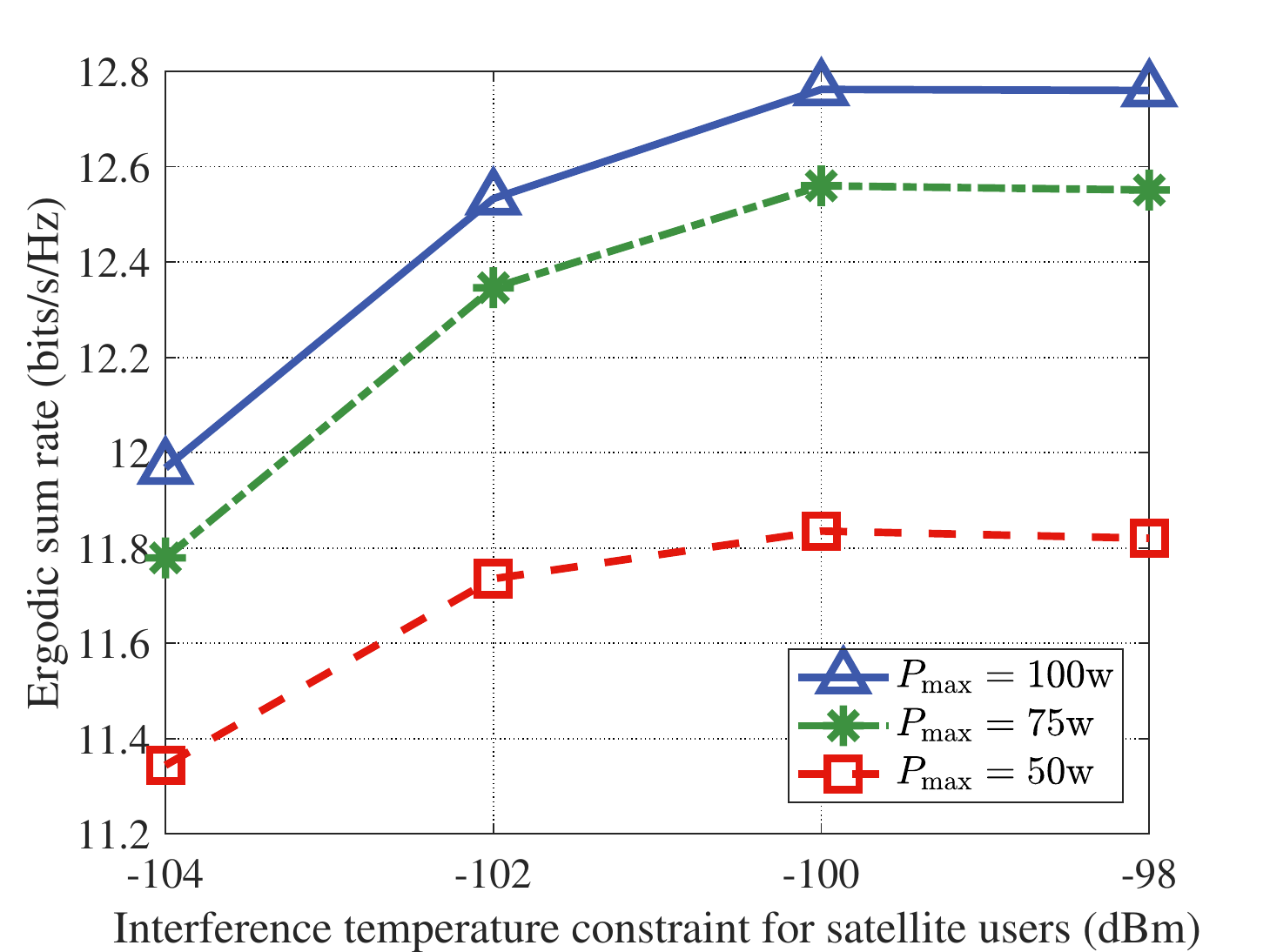}
	\caption{The ergodic sum rate with  different transmit power constraints, varying with the interference constraint of satellite users. }
	\label{fig3}
	\end{figure}

	\begin{figure}
	\centering
	\includegraphics[width=3.3 in]{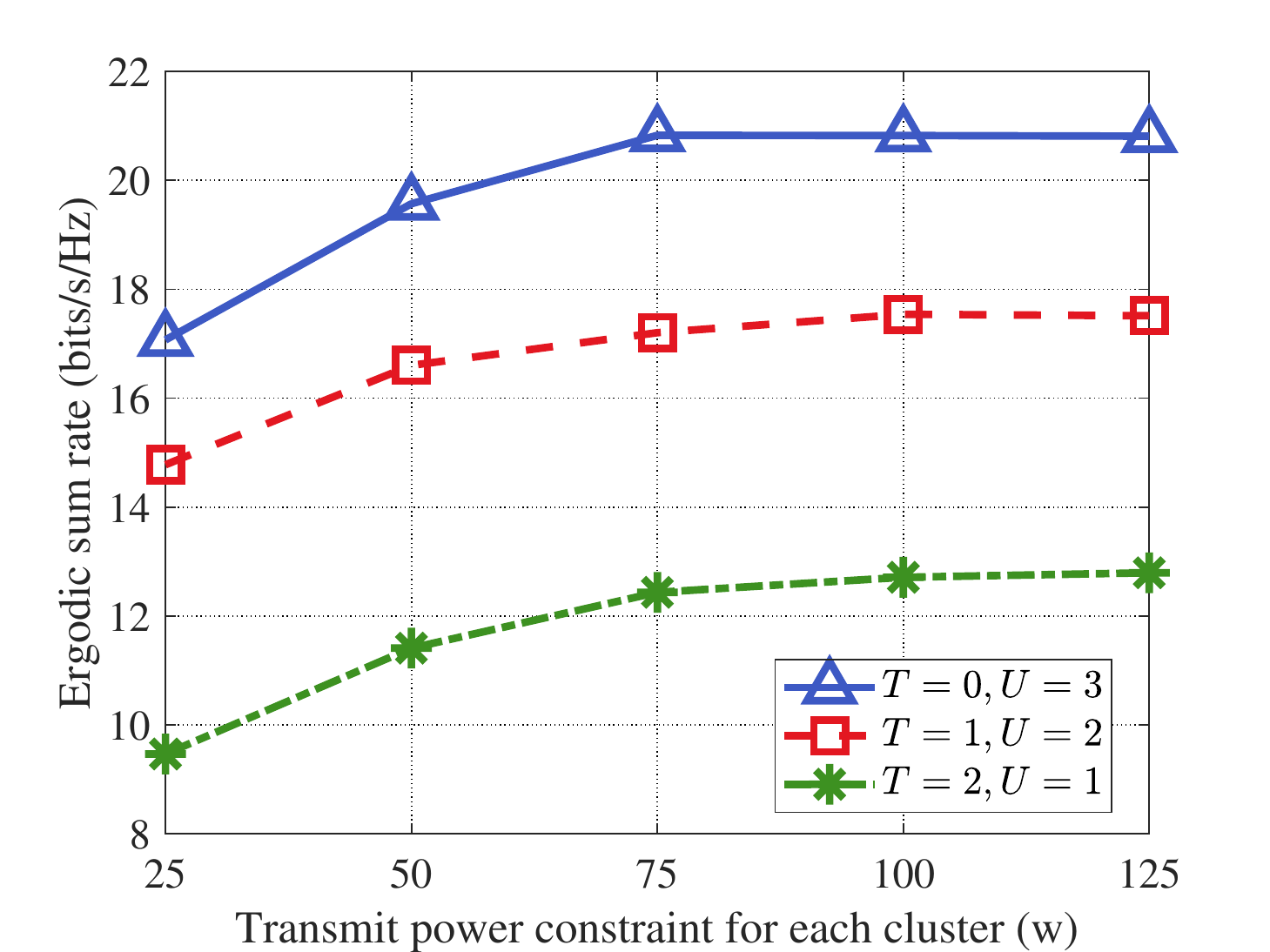}
	\caption{The ergodic sum rate with different configurations of TBSs and UAVs.}
	\label{fig5}
	\end{figure}

	Secondly, we compare the proposed scheme based on NOMA with three OMA schemes proposed in \cite{ref3} and \cite{ref5} as well as the scheme of equal power allocation. Without loss of generality, the allocation coefficient of two users is set to be $\frac{1}{2}$\footnote{
	Since the scheme proposed in  \cite{ref3} and  \cite{ref5} did not consider the minimum user rate constraint, we  set $r_k=0$ bits/s/Hz $(k=1,...,K)$ in this simulation for comparisons.}. As shown in Fig. \ref{fig2}, it is obvious that the proposed NOMA scheme outperforms remarkably the other three OMA schemes. These gains are achieved at the cost of more complex receivers. The inter-user interference is reduced by SIC. In addition, since NOMA separates users in the power domain, it provides additional flexibility to assign resources between paired users. But for OMA, it is hard to achieve the flexible time or frequency partition so that the resource efficiency is not fully exploited. Therefore, based on the SIC reception and effective resource allocation, NOMA is more capable to provide on-demand services for different users. In addition, one can also find that the scheme proposed in \cite{ref3} performs better than the other two OMA schemes. It can be explained by the fact that in \cite{ref3}, the interference is effectively controlled based on the large-scale CSI. This further shows that positions are critical factors that can be utilized to mitigate co-channel interference for irregular settings.

	Next, we evaluate the impact of different rate requirements of user 2. As we can see from Fig. \ref{fig4}, the system experiences very small performance degradation as the rate requirement increases, only less than 6\% in our settings. It proves that NOMA could greatly support the wide sparsity of maritime IoT, satisfying the basic QoS requirement of user 2 without sacrificing much system efficiency to provide high-rate connections for user 1.

	The impact of inter-segment interference is shown in Fig. \ref{fig3}. In the beginning, the sum rate increases with the interference constraint. The nearshore network has to control its power to ensure the normal operations of satellite receivers. But when the interference constraint is relaxed more than $-100$dBm, the sum rate maintains unchanged. In this situation, the nearshore network and marine satellites can coexist without any compromise.

	In Fig.~\ref{fig5}, we show the superiority of UAVs for the coverage enhancement. Different from the above settings, in this simulation, we keep the topology unchanged and replace the original TBSs with UAVs or vice versa. We set $T_1=T_2=T_3=T, \ U_1=U_2=U_3=U$. Obviously, UAVs bring considerable system gains compared with TBSs, with more than 50\% rate improvement of one more UAV and one less TBS in our settings. This is mainly because of UAVs' height merit which helps build high-quality LOS links. Besides, we want to note that the case of only TBSs is not presented for the fact that only TBSs cannot support the basic need of user 2. This indicates that with the assistance of UAVs, the coverage scope of the nearshore network is enlarged. 

	\begin{figure}[t]
	\centering
	\includegraphics[width=3.3 in]{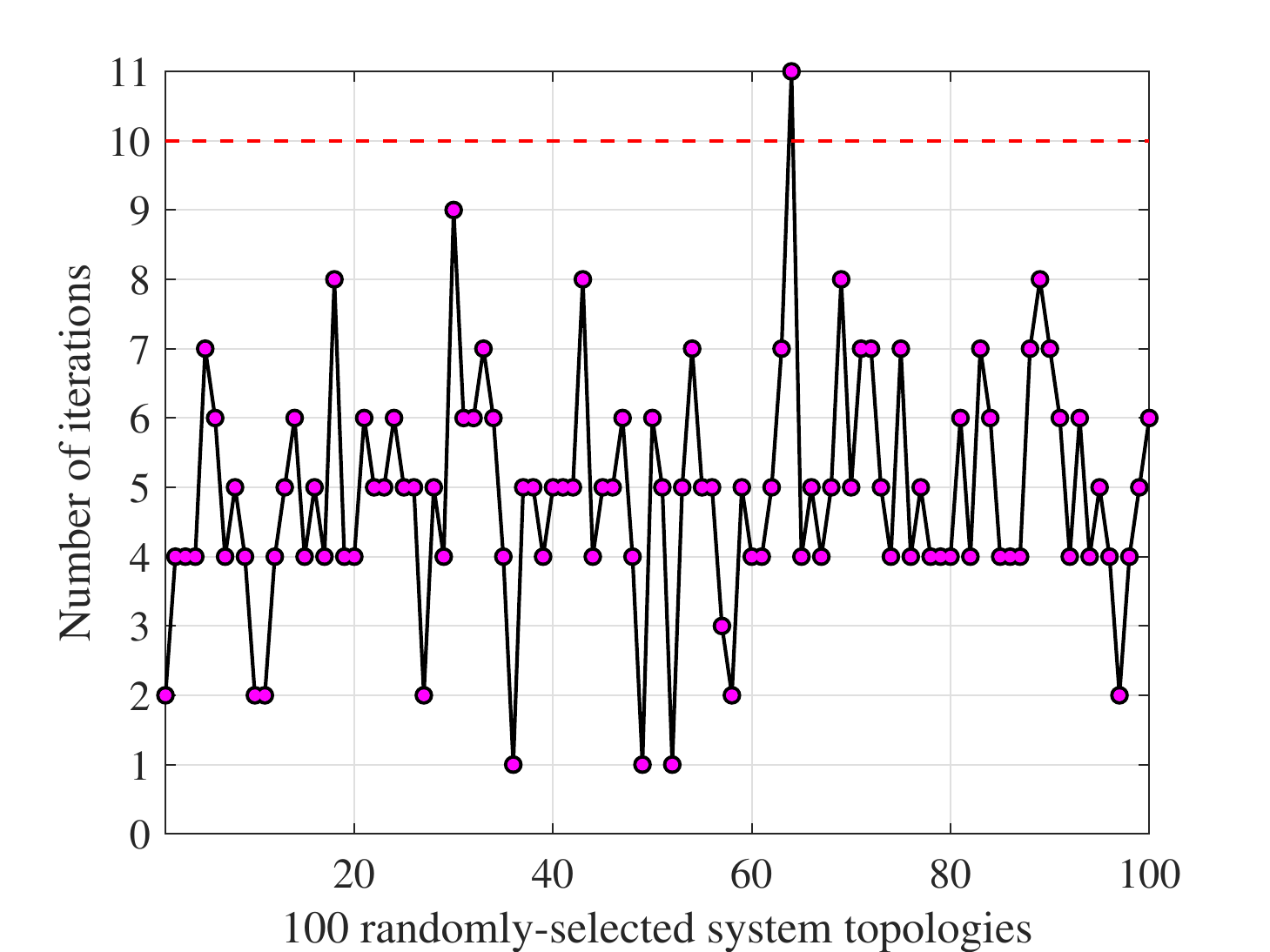}
	\caption{Convergence performance of the proposed iterative algorithm.}
	\label{fig6}
\end{figure}

Finally, the algorithm convergence is shown in Fig. \ref{fig6}. We randomly generate 100 different typologies with $P_\text{max}=50$w. In most cases, the proposed iterative algorithm only takes less than 10 times to converge, which shows great potentials to achieve timely resource allocation in reality.

\section{Conclusions}
\label{section 5}
In this paper, we have investigated a NOMA-based hybrid satellite-UAV-terrestrial network on the sea. Particularly, we have utilized  tethered UAVs to coordinate with TBSs. They form virtual clusters and serve users in a user-centric manner. To accommodate the wide sparsity of maritime IoT, the NOMA technique has been applied to agilely serve different users. The spectrum sharing has been considered between nearshore clusters and  marine satellites. We have proposed a joint power allocation scheme to tackle the challenging interference among different users, different clusters, and different network segments. Based on large-scale CSI only, the ergodic sum rate has been maximized at a low cost. To tackle the non-convex problem, we transformed our problem into a max-min problem and constructed an auxiliary function to simplify the troubling non-linear constraints. An iterative algorithm has been proposed to find the solution by solving a series of optimization-friendly subproblems. Simulation results have shown the potentials of NOMA-based hybrid satellite-UAV-terrestrial networks for maritime coverage enhancement.

\appendices
\section{Proof of \textit{Theorem 1}}
	Through observations,  \eqref{eqd3} can be rewritten as 
\begin{equation}
	\label{ad5}
	\begin{aligned}
		&1-(v^*_{k,2})^{-1}=\sum_{n=1}^{T_k+U_k}\frac{l_{2,n}^{(k,k)}P_{2,n}^{(k)}}{v^*_{k,2}\sigma^2_{k,2}(\mathbf{P})+M_kl_{2,n}^{(k,k)}P_{2,n}^{(k)}}.
	\end{aligned}
\end{equation}
Based on this transformation, the term $1-(v^*_{k,2})^{-1}$ in $g_{k,2}(\mathbf{P},v_{k,2}^*)$ (defined in \eqref{eq20}) can be replaced by the right side of \eqref{ad5}. Consequently, a new function $f_{k}(\mathbf{P},z_k)$ is obtained
\begin{equation}
	\begin{aligned}
		f_{k}(\mathbf{P},z_k)
		=\sum\limits_{n=1}^{T_k+U_k}\bigg(\log_2\big(1+\frac{M_kl_{2,n}^{(k,k)}P_{2,n}^{(k)}}{z_k\sigma^2_{k,2}(\mathbf{P})}\big)&\\				-\frac{\log_2(e)M_kl_{2,n}^{(k,k)}P_{2,n}^{(k)}}{z_k\sigma^2_{k,2}(\mathbf{P})+M_kl_{2,n}^{(k,k)}P_{2,n}^{(k)}}\bigg)+M_k\log_2z_k,&
	\end{aligned}
\end{equation}
where $z_k$ is the introduced variable corresponding to $v_{2,k}^*$ and we have 
\begin{equation}
	\label{ad6}
	g_{k,2}(\mathbf{P},v_{2,k}^*)=f_{k}(\mathbf{P},v_{2,k}^*).
\end{equation}
Next, we proof the monotonicity of $f_{k}(\mathbf{P},z_{k})$ with respect to $z_{k}$. We calculate the partial derivative of $f_{k}(\mathbf{P},z_{k})$,
\begin{equation}
	\begin{aligned}
		&\frac{\partial f_{k}(\mathbf{P},z_{k})}{\partial z_{k}}=\\
		&\frac{\log_2e}{z_{k}}\bigg[M_k-\sum_{n=1}^{T_k+U_k}\big(\frac{M_kl_{2,n}^{(k,k)}P_{2,n}^{(k)}}{z_{k}\sigma^2_{k,2}(\mathbf{P})+M_kl_{2,n}^{(k,k)}P_{2,n}^{(k)}}\big)^2\bigg]\\
		&> \frac{\log_2e}{z_{k}}[M_k-(T_k+U_k)].
	\end{aligned}
\end{equation}
When $M_k\geqslant T_k+U_k$, the  partial derivative is positive.
It indicates that  $f_{k}(\mathbf{P},z_{k})$  is monotonically increasing with $z_{k}$. Based on the monotonicity of $f_{k}(\mathbf{P},z_{k})$ and the equation \eqref{ad6}, we can express $g_{k,2}(\mathbf{P},v_{k,2}^*)$ in an optimization form as follows
\begin{equation}
	g_{k,2}(\mathbf{P},v_{k,2}^*)=\max\limits_{1\leqslant z_k\leqslant v_{k,2}^*}f_{k}(\mathbf{P},z_k),
\end{equation}
and \textit{Theorem 1} is proved.

\section{Proof of \textit{Lemma 1}}
We prove this by using the method of reduction to absurdity.
First, if $\{\hat{\mathbf{P}}, \hat{z}_k,\hat{\mathbf{t}}_k, \hat{\mathbf{v}}\}$ is an optimal solution to (P5), and thus $\hat{\mathbf{P}}$ is also the optimal solution to (P2). 
Assuming $\hat{\mathbf{P}}$ is not the optimal solution to (P2), we could find another solution, defined as $\tilde{\mathbf{P}}$, which is optimal to (P2) and we have
\begin{equation}
	g(\tilde{\mathbf{P}},\tilde{\mathbf{v}}) > g(\hat{\mathbf{P}},\hat{\mathbf{v}}),
\end{equation}
where  $\hat{\mathbf{v}}$ and $\tilde{\mathbf{v}}$ are the minimal point of $g(\hat{\mathbf{P}},\mathbf{v})$ and $g(\tilde{\mathbf{P}},\mathbf{v})$, which is uniquely determined  by $\hat{\mathbf{P}}$ and $\tilde{\mathbf{P}}$ as shown in \eqref{eqd1}.
Based on $\tilde{\mathbf{P}}$, we define $\tilde{z}_k$ and $\tilde{\mathbf{t}}_k=\{\tilde{t}_{k,n}| n=1,...,T_k+U_k\},  k=1,...,K$ as follows
\begin{subequations}	
	\begin{align}
		\tilde{z}_k&=\tilde{v}_{k,2},	 \\
		\tilde{t}_{k,n}&=\frac{M_kl_{2,n}^{(k,k)}\tilde{P}_{2,n}^{(k)}}{\tilde{z}_k\sigma^2_{k,2}(\tilde{\mathbf{P}})+M_kl_{2,n}^{(k,k)}\tilde{P}_{2,n}^{(k)}}. \label{ad12}
	\end{align}
\end{subequations}
According to \eqref{eqn31}, we could derive that
\begin{subequations}
	
	\begin{align}
		\tilde{z}_k&=\tilde{v}_{k,2}
		\\
	\Rightarrow	&(\tilde{z}_k)^{-1}+\sum_{n=1}^{T_k+U_k}\frac{l_{2,n}^{(k,k)}\tilde{P}_{2,n}^{(k)}}{\tilde{z}_k\sigma^2_{k,2}(\tilde{\mathbf{P}})+M_kl_{2,n}^{(k,k)}\tilde{P}_{2,n}^{(k)}} = 1 \\
	\Rightarrow	&(\tilde{z}_k)^{-1}+\frac{1}{M_k}\sum_{n=1}^{T_k+U_k}\tilde{t}_{k,n}=1. \label{ad11}
	\end{align}
\end{subequations}
Referring to \eqref{eq30}, \eqref{eqn30} and \eqref{ad8}, there exist following equations
\begin{equation}
	g_{k,2}(\tilde{\mathbf{P}},\tilde{v}_{k,2})=f_{k}(\tilde{\mathbf{P}},\tilde{v}_{k,2})=f_{k}(\tilde{\mathbf{P}},\tilde{z}_{k})=\mathcal{F}_k(\tilde{z}_k,\tilde{\mathbf{t}}_k).
\end{equation}
Since $\tilde{\mathbf{P}}$ is optimal to (P2), it sure meets the constraint \eqref{eq4}, so, we could deduce that
\begin{equation}
	\label{ad9}
	g_{k,2}(\tilde{\mathbf{P}},\tilde{v}_{k,2}) \geqslant r_k \Rightarrow  \mathcal{F}(\tilde{z}_k,\tilde{\mathbf{t}}_k) \geqslant r.
\end{equation}
As shown in \eqref{ad12}, \eqref{ad11} and \eqref{ad9}, we could find $\{\tilde{\mathbf{P}}, \tilde{z}_k, \tilde{\mathbf{t}}_k, \tilde{\mathbf{v}}\}$ is feasible to (P5)
and we have $g(\tilde{\mathbf{P}},\tilde{\mathbf{v}}) > g(\hat{\mathbf{P}},\hat{\mathbf{v}})$, which is equivalent to 
\begin{equation}
	\min\limits_{\mathbf{v}} g(\tilde{\mathbf{P}},\mathbf{v}) > \min\limits_{\mathbf{v}} g(\hat{\mathbf{P}},{\mathbf{v}}).
\end{equation}
It is contradictory that $\{\hat{\mathbf{P}}, \hat{z}_k,\hat{\mathbf{t}}_k, \hat{\mathbf{v}}\}$ is the optimal solution to (P5). So the assumption that there exists 
other optimal solution to (P2) does not hold and $\hat{\mathbf{P}}$ is optimal to (P2).

Similarly, if $\tilde{\mathbf{P}}$ is  the optimal solution to (P2), it also holds the optimality in (P5). If not, there exists $\{\hat{\mathbf{P}}, \hat{z}_k,\hat{\mathbf{t}}_k, \hat{\mathbf{v}}\}$ and we have 
\begin{subequations}
	\begin{align}
		\min\limits_{\mathbf{v}} g(\hat{\mathbf{P}},{\mathbf{v}})&>\min\limits_{\mathbf{v}} g(\tilde{\mathbf{P}},\mathbf{v}), 	\\
		\Rightarrow g(\hat{\mathbf{P}},\hat{\mathbf{v}})&> g(\tilde{\mathbf{P}},\tilde{\mathbf{v}}).
	\end{align}
\end{subequations}
Based on $\hat{\mathbf{P}}$, define $\hat{z'}_k$ and $\hat{\mathbf{t'}}_k$ as follows
\begin{subequations}	
	\begin{align}
		\hat{z'}_k&=\hat{v}_{k,2},	 \\
		\hat{t'}_{k,n}&=\frac{M_kl_{2,n}^{(k,k)}\hat{P}_{2,n}^{(k)}}{\hat{z'}_k\sigma^2_{k,2}(\hat{\mathbf{P}})+M_kl_{2,n}^{(k,k)}\hat{P}_{2,n}^{(k)}}.
	\end{align}
\end{subequations}
Referring to \eqref{ad3} and \eqref{ad4}, it is easy to find that $\hat{z'}_k\geqslant \hat{z}_k$ and $\hat{t'}_{k,n}\geqslant\hat{t}_{k,n}$. Based on the  monotonically increasing characteristic of $\mathcal{F}_k(z_k,\mathbf{t}_k)$, we have that
\begin{equation}
	\mathcal{F}_k(\hat{z'}_k,\hat{\mathbf{t'}}_k) \geqslant \mathcal{F}_k(\hat{z}_k,\hat{\mathbf{t}}_k).
\end{equation}
According to \eqref{eq30}, \eqref{eqn30} and \eqref{ad8}, the following equations  hold
\begin{equation}
	g_{k,2}(\hat{\mathbf{P}},\hat{v}_{k,2})=f_{k}(\hat{\mathbf{P}},\hat{v}_{k,2})=f_{k}(\hat{\mathbf{P}},\hat{z'}_k)=\mathcal{F}(\hat{z'}_k,\hat{\mathbf{t'}}_k). 
\end{equation}
Since $\mathcal{F}_k(\hat{z}_k,\hat{\mathbf{t}}_k) \geqslant r_k$, it is affirmed that $\mathcal{F}_k(\hat{z'}_k,\hat{\mathbf{t'}}_k)   \geqslant r_k$ and thus $g_{k,2}(\hat{\mathbf{P}},\hat{v}_{k,2}) \geqslant r_k$. Consequently, $\hat{\mathbf{P}}$ is a feasible solution to (P2) and it achieves higher value than $\tilde{\mathbf{P}}$, which is contradictory that $\tilde{\mathbf{P}}$ is the optimal solution to (P2). So, the assumption that there exists other optimal solution to (P5) does not hold. The optimal equivalence between (P2) and (P5) is thus proved.

\section{Proof of \textit{Lemma 2}}
By applying the Taylor expansion, the tangent of the concave function is above its curve and the tangent of the convex function is blow its curve. Since $\log_2(1-t_{k,n})$ (the term in $\mathcal{F}_k(z_k,\mathbf{t}_k)$) and $z_k^{-1}$ are convex with respect to $t_{k,n}$ and $z_k$, $\mbox{In}(z_k)+\mbox{In}(t_{k,n})+\mbox{In}({\sigma}^{2}_{k,2}(\mathbf{P}))$ is concave with respect to $z_k, \ t_{k,n}$ and $\mathbf{P}$, the following inequalities hold
\begin{subequations}
	\begin{equation}
		\mathcal{F}_k(z_k,\mathbf{t}_k) \geqslant 	\mathcal{F}_k(z_k,\mathbf{t}_k|\mathbf{t}_k^{s-1}),
	\end{equation}
	\begin{equation}
		\frac{1}{z_k}\geqslant-\frac{z_k}{(z_k^{s-1})^2}+\frac{1}{z_k^{s-1}},
	\end{equation}
	\begin{equation}
		\begin{aligned}
			\mbox{In}(z_k)+\mbox{In}(t_{k,n})+&\mbox{In}({\sigma}^{2}_{k,2}(\mathbf{P})) 
			\leqslant \\
			\mbox{In}(z_k|z_k^{s-1})+&\mbox{In}(t_{k,n}|t_{k,n}^{s-1})+\mbox{In}(\sigma^2_{k,2}(\mathbf{P}|\mathbf{P}^{s-1})),
		\end{aligned}
	\end{equation}
\end{subequations}
which make the constraints \eqref{eq40}, \eqref{eq41} and \eqref{eq42} in (P7) are tighter than the corresponding constraints in (P6).  Thereby, in each iteration, the optimal solution $\mathbf{P}^s$ is guaranteed in the feasible set of the original problem.

As for the objection, since $\mathcal{G}_{k,m,2}(\mathbf{P})$ is a concave function with respect to $\mathbf{P}$, it holds that
\begin{equation}
	\mathcal{G}_{k,m,2}(\mathbf{P}|\mathbf{P}^{s-1}) \geqslant \mathcal{G}_{k,m,2}(\mathbf{P}).
\end{equation}
Accordingly, the following inequalities can be obtained
\begin{equation}
	\begin{aligned}
		\mathcal{G}_{k,m}(\mathbf{P}, w_{k,m})&=\mathcal{G}_{k,m,1}(\mathbf{P}, w_{k,m})-\mathcal{G}_{k,m,2}(\mathbf{P})\\
		&\geqslant \mathcal{G}_{k,m,1}(\mathbf{P}, w_{k,m}) -\mathcal{G}_{k,m,2}(\mathbf{P}|\mathbf{P}^{s-1}) \\
		&=\mathcal{G}_{k,m}(\mathbf{P}, w_{k,m}|\mathbf{P}^{s-1}).
	\end{aligned}
\end{equation}
So, we have that
\begin{equation}
	\label{ap2}
	\begin{aligned}
		\mathcal{G}(\mathbf{P}, \mathbf{w})\geqslant \mathcal{G}(\mathbf{P}, \mathbf{w}|\mathbf{P}^{s-1}).
	\end{aligned}
\end{equation}
Furthermore, according to the properties of the saddle point, there are following conclusions  for $(\mathbf{P}^s,\mathbf{w}^s)$
\begin{equation}
	\label{ap1}
	\begin{aligned}
		&\mathcal{G}(\mathbf{P}^s, \mathbf{w}^s|\mathbf{P}^{s-1})\geqslant \mathcal{G}(\mathbf{P}, \mathbf{w}^s|\mathbf{P}^{s-1}), \\
		&\mathcal{G}(\mathbf{P}^s, \mathbf{w}^s|\mathbf{P}^{s-1})\leqslant \mathcal{G}(\mathbf{P}^s, \mathbf{w}|\mathbf{P}^{s-1}).
	\end{aligned}
\end{equation}
Based on \eqref{ap2} and \eqref{ap1}, the relationship of $\mathcal{G}(\mathbf{P}^s, \mathbf{w}^s)$ and $\mathcal{G}(\mathbf{P}^{s-1}, \mathbf{w}^{s-1})$ can be derived as follows
\begin{subequations}
	\begin{align}
		\mathcal{G}(\mathbf{P}^s, \mathbf{w}^s)&\geqslant \mathcal{G}(\mathbf{P}^s, \mathbf{w}^s|\mathbf{P}^{s-1})=\max\limits_{\mathbf{P}} g(\mathbf{P}, \mathbf{w}^s|\mathbf{P}^{s-1})\\
		&\geqslant \mathcal{G}(\mathbf{P}^{s-1}, \mathbf{w}^s|\mathbf{P}^{s-1})=\mathcal{G}(\mathbf{P}^{s-1}, \mathbf{w}^s)\\
		&\geqslant \min\limits_{\mathbf{w}} \mathcal{G}(\mathbf{P}^{s-1}, \mathbf{w})=\mathcal{G}(\mathbf{P}^{s-1}, \mathbf{w}^{s-1}).
	\end{align}
\end{subequations}
Based on the monotonically increasing property of the objective function in iterations and the fact that the ergodic sum rate is up-bounded, the proposed algorithm is assured to converge, at least to a local optimum.
\end{document}